\documentclass[11pt]{article}
\usepackage{jheppub}
\usepackage{amsmath,amssymb,amsfonts,graphicx}

\newcommand{\be}{\begin{equation}}
\newcommand{\ee}{\end{equation}}
\newcommand{\bea}{\begin{eqnarray}}
\newcommand{\eea}{\end{eqnarray}}
\newcommand{\beas}{\begin{eqnarray*}}
\newcommand{\eeas}{\end{eqnarray*}}
\newcommand{\ba}{\begin{array}}
\newcommand{\ea}{\end{array}}

\renewcommand*\d[2][]{%
	\mathrm{d}%
	\ifx\relax#1\relax\else
	\rule{-0.02em}{1.5ex}^{#1}\rule{0.08em}{0ex}\!
	\fi
	#2\,
}

\title{Cosmology from confinement?}

\author[]{Mark Van Raamsdonk}

\affiliation[]{Department of Physics and Astronomy, University of British Columbia,\\
6224 Agricultural Road, Vancouver, B.C.\ V6T 1Z1, Canada.}

\emailAdd{mav@phas.ubc.ca}

\abstract{We describe a class of holographic models that may describe the physics of certain four-dimensional big-bang / big-crunch cosmologies. The construction involves a pair of 3D Euclidean holographic CFTs each on a homogeneous and isotropic space $M$ coupled at either end of an interval ${\cal I}$ to a Euclidean 4D CFT on $M \times {\cal I}$ with many fewer local degrees of freedom. We argue that in some cases, when the size of $M$ is much greater than the length of ${\cal I}$, the theory flows to a confining three-dimensional field theory on $M$ in the infrared, and this is reflected in the dual description by the asymptotically AdS spacetimes dual to the two 3D CFTs joining up in the IR to give a Euclidean wormhole. The Euclidean construction can be reinterpreted as generating a state of the Lorentzian 4D CFT on $M \times {\rm time}$ whose dual includes the physics of a big-bang / big-crunch cosmology. When $M$ is $\mathbb{R}^3$, we can alternatively analytically continue one of the $\mathbb{R}^3$ directions to get an eternally traversable four-dimensional planar wormhole. We suggest explicit microscopic examples where the 4D CFT is ${\cal N}=4$ SYM theory and the 3D CFTs are superconformal field theories with opposite orientation. In this case, the two geometries dual to the pair of 3D SCFTs can be understood as a geometrical version of a brane-antibrane pair, and the tendency of the geometries to connect up is related to the standard instability of brane-antibrane systems.}

\notoc

\begin{document}

\maketitle
\newpage
\parskip=10pt

\section{Introduction}

In this note we describe specific holographic constructions through which the physics of 4D big-bang/big-crunch cosmologies might be encoded in the physics of certain non-gravitational quantum field theories. We follow the construction of \cite{Cooper2018,Antonini2019,VanRaamsdonk:2020tlr} (reviewed below; see also \cite{Penington:2019kki, Dong:2020uxp, Chen:2020tes} for related low-dimensional constructions), which considers states of a 4D holographic CFT constructed using a Euclidean BCFT path integral. These states were suggested to be dual to asymptotically AdS black hole spacetimes with a dynamical end-of-the-world (ETW) brane providing an inner boundary for the spacetime behind the horizon of the black hole. In \cite{Cooper2018} was suggested that in favorable cases, gravity can localize to this ETW brane, so that the effective description of the ETW brane physics is that of a four-dimensional big-bang/big-crunch cosmology.

The present work refines and extends this picture in the following ways:
\begin{itemize}
\item
We point out that the physics of confinement and symmetry breaking plays a crucial role in the relevant field theories.\footnote{Related comments were made in the context of Euclidean wormholes in \cite{Betzios:2019rds}.}
The construction relies on four-dimensional field theories with one compact direction flowing in the infrared to three-dimensional confined theories, with a particular pattern of global symmetry breaking.
\item
We suggest specific microscopic examples constructed from ${\cal N}=4$ SYM theory and various 3D superconformal field theories that can be coupled to it at a boundary. The dual gravity picture, including the ETW brane physics, is described in terms of type IIB supergravity / string theory, and involves the physics of brane-antibrane systems.
\item
We emphasize that the construction may continue to work when the 4D theory is not a conventional holographic theory (but the boundary theories involved in constructing the state are). In this case, there can be a classical gravitational description for the ETW brane (as a 4D theory of gravity, perhaps with a compact internal space) but no classical bulk 5D spacetime. The encoding of a cosmological spacetime in the state of a non-holographic CFT is similar to the encoding of black hole interiors in Hawking radiation systems \cite{Maldacena:2013xja, Penington:2019npb, Almheiri:2019psf}.
\end{itemize}

While we don't attempt to construct the relevant supergravity solutions in detail, we are able to describe the asymptotic behaviour explicitly. It remains to check (or argue indirectly) that the proposed solutions exist and have the conjectured properties. Alternatively, we can hope to understand better the 4D effective description of the ETW physics and verify that the desired solutions relevant to cosmology exist there. If the construction succeeds, an analytically-continued version gives four-dimensional eternally traversable wormholes preserving 2+1 dimensional Poincar\'e symmetry in the effective description. It has been argued that the existence of these would require an unnaturally large amount of negative null energy \cite{Freivogel:2019lej,LinMaldacena}. We review these arguments in Section \ref{sec:effective} and identify a novel field theory effect that gives a possible mechanism for achieving the large amount of negative null energy required to support the wormhole.

These models have some interesting phenomonological/model-building aspects that we discuss in Section \ref{sec:discussion}.
However, we emphasize that the immediate motivation here is not to come up with a phenomenologically accurate model of cosmology, but rather to come up with {\it some} completely defined physical theory which encodes the cosmological physics of a four-dimensional homogeneous and isotropic universe with a big bang.\footnote{For other approaches to cosmology using holography, see for example \cite{Banks:2001px, Strominger:2001pn, Alishahiha:2004md, Freivogel:2005qh, McFadden:2009fg}.} If these constructions succeed, they could shed light on the question of what are the well-defined observables in cosmological spacetimes and allow a first principles calculation of these observables assuming the holographic dictionary is understood well enough and the field theory calculations can be done. For example, cosmological correlators could be computed from correlation functions in a four-dimensional Euclidean field theory with boundaries at some past and future Euclidean time.

\subsubsection*{Summary of the basic construction}

\begin{figure}
\centering
\includegraphics[width=80mm]{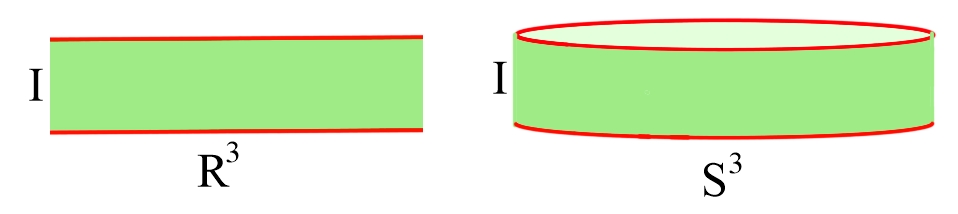}
\caption{Basic field theory construction (the CFT sandwich): a pair of 3D holographic CFTs related by a reflection are coupled at either end of an interval $I$ to a 4D CFT.}
\label{fig:basic}
\end{figure}

We begin by describing the basic mechanism of the construction, considering the case where we wish to describe cosmology with spatial geometry $\mathbb{R}^3$. The construction is essentially the same when the spatial geometry is spherical or hyperbolic. Following \cite{Maldacena:2004rf}, we begin by constructing a Euclidean wormhole.

To start, consider a pair of three-dimensional Euclidean holographic CFTs each living on $\mathbb{R}^3$. Each of these is dual to a separate four-dimensional Euclidean gravitational theory on AdS${}^4$ with boundary geometry $\mathbb{R}^3$. We now introduce an interaction between the theories by coupling them to an auxiliary four-dimensional quantum field theory on $\mathbb{R}^3$ times an interval $[- \tau_0/2,  \tau_0/2]$, with the original three-dimensional theories living at either end of the interval, as in Figure \ref{fig:basic}a.\footnote{In other words, we make a CFT sandwich.} We take the four-dimensional theory to have many fewer local degrees of freedom than the original 3D theories. In particular, the four-dimensional theory need not be a conventional holographic theory. Since the fourth dimension is compact, the field theory we have constructed will flow to some three-dimensional theory in the IR, which provides a good description of the physics at distance scales much larger than $\tau_0$. This theory could be a non-trivial three-dimensional conformal field theory, but more generically, we expect that it will be gapped/confining. We will argue that in some cases, the gravitational description of this confinement is that the asymptotically AdS${}^4$ spacetimes associated with the two 3D CFTs join up in the IR so that the full spacetime is a Euclidean wormhole.

\begin{figure}
\centering
\includegraphics[width=140mm]{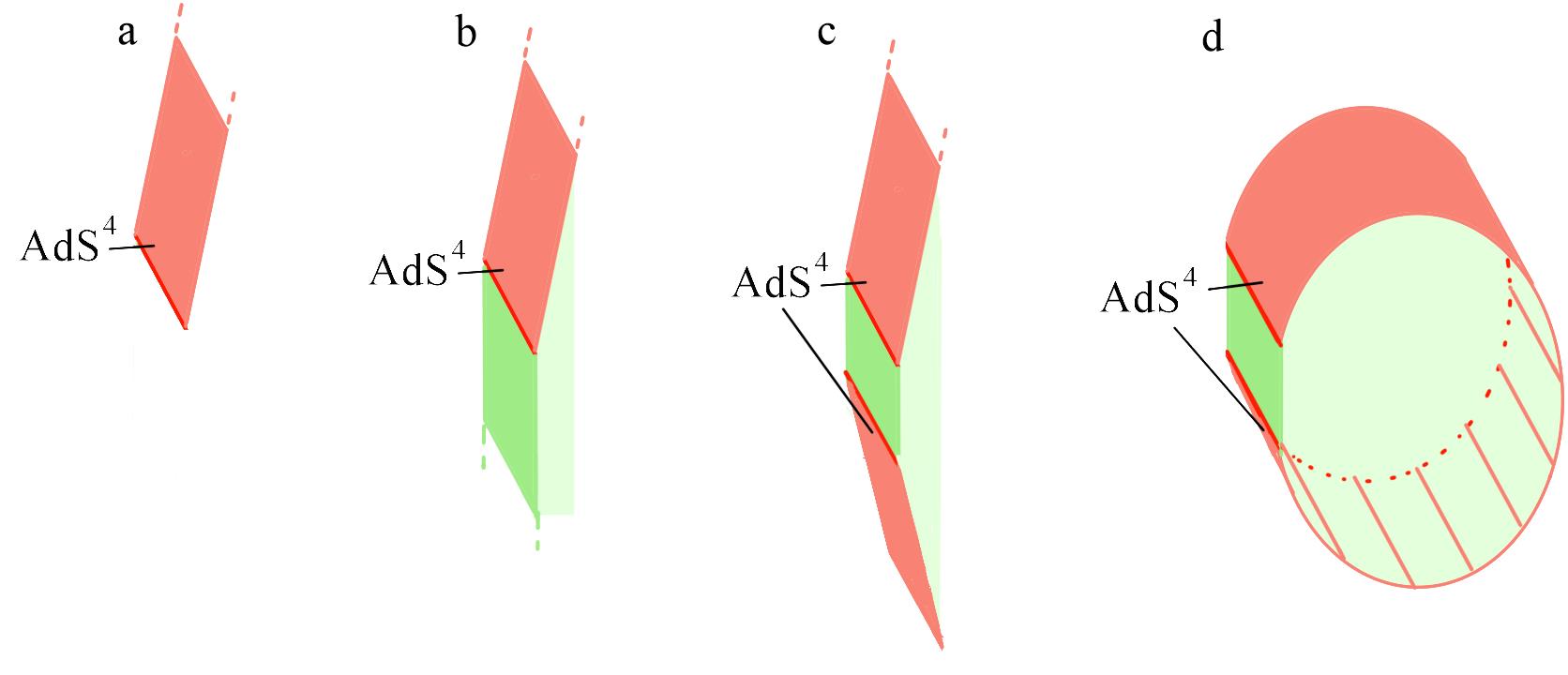}
\caption{Dual geometries for various field theory setups, showing end-of-the-world branes (red) with asymptotically AdS${}^4$ regions. (a) Dual of a single 3D CFT (b) Dual of a 3D CFT coupled to the boundary of a 4D CFT. Gravity remains well-localized to the ETW brane when $c_{4D} \ll c_{3D}$. (c) Possible dual of a pair of 3D CFTs coupled to a 4D CFT, where the IR physics is a conformal 3D CFT (d) Possible dual of a pair of 3D CFTs coupled to a 4D CFT, where the IR physics is a confining 3D theory.}
\label{fig:dual}
\end{figure}

To motivate this assertion, consider the case where the four-dimensional auxiliary theory is also holographic. In this case, coupling one of the 3D holographic theories to the auxiliary 4D system gives a holographic boundary conformal field theory. In the dual description, the four-dimensional gravitational theory dual to the 3D CFT now describes the physics of an end-of-the-world brane in a five-dimensional geometry (Figure \ref{fig:dual}b). When the 4D theory has many fewer local degrees of freedom than the 3D theory ($c_{4D} \ll c_{3D}$), gravity localizes to this ETW brane by the Karch-Randall mechanism \cite{Karch:2000ct,Karch:2000gx,Karch:2001cw}, and the 4D graviton gets a mass that can be made arbitrarily small by taking $c_{4D}/c_{3D}$ small. In our construction with two 3D theories on either end of an interval, we have two ETW branes in the UV. But if the full field theory is confining in the IR, the dual geometry must be capped off somehow in the IR \cite{Witten1998a}, and a natural mechanism for this is for the two ETW branes to join up (Figure \ref{fig:dual}d).\footnote{As we discuss below, there are more general possibilities for which the effective description of the ETW brane physics is not a single 4D Euclidean wormhole.} We will provide evidence for this picture via a string theory construction, where the 4D auxiliary theory is taken to be ${\cal N}=4$ SYM theory and the 3D theories are holographic superconformal theories with opposite orientation. In this case, the ETW branes are related to a certain brane-antibrane system in string theory, and the tendency for the ETW branes to join up is directly related to the instability of the brane-antibrane system. From the field theory perspective, this situation is characterized by a spontaneous breaking $G \times G \to G$ of global symmetry, where $G$ is the global symmetry associated with each of the 3D CFTs and becomes a gauge group for gauge fields on the ETW brane.

\begin{figure}
\centering
\includegraphics[width=140mm]{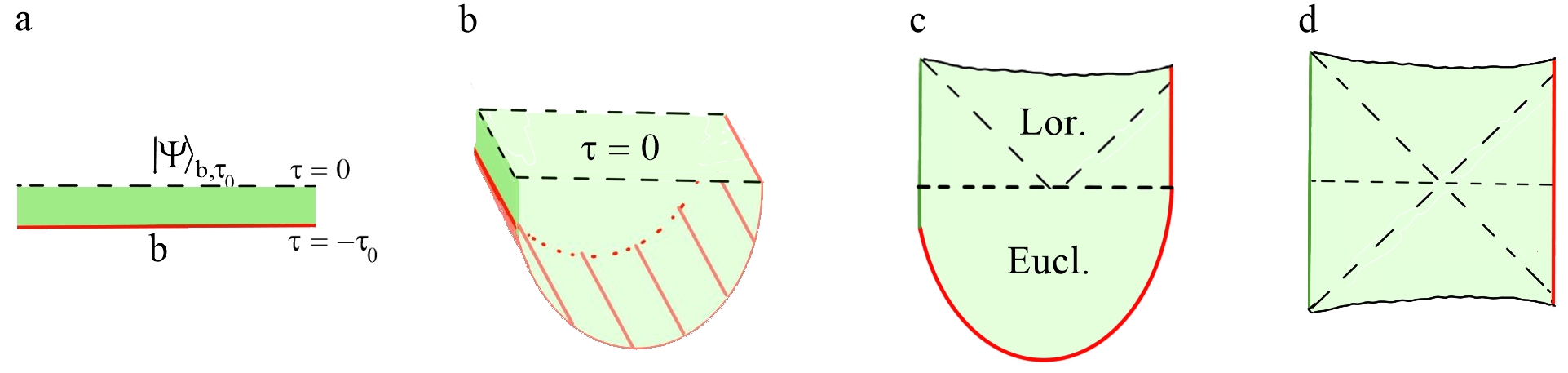}
\caption{Connection to cosmology. (a) State of the 4D CFT on $\mathbb{R}^3$ produced by the Euclidean path integral terminated by a 3D CFT $b$ in the Euclidean past at $\tau = - \tau_0$. (b) $\tau < 0$ half of the Euclidean solution dual to the doubled bra-ket path-integral. (c) The $\tau = 0$ slice of the Euclidean solution serves as the initial data for Lorentzian evolution. (d) Full Lorentzian solution dual to $|\Psi\rangle_{b, \tau_0}$. }
\label{fig:cosmo}
\end{figure}

To connect with cosmology, we interpret the $\tau$ direction as a Euclidean time direction and interpret the Euclidean theory for $\tau < 0$ as a path-integral that constructs a specific state $|\Psi_{b,\tau_0}\rangle$ of our auxiliary 4D theory living on a spatial $\mathbb{R}^3$ (Figure \ref{fig:cosmo}a).\footnote{This is similar to the Hartle-Hawking construction \cite{Hartle:1983ai}, but with a CFT path integral.} Here, $b$ labels our choice of 3D theory. Note that the degrees of freedom of this 3D theory are not physical degrees of freedom in the Lorentzian theory, but appear only in the Euclidean path integral generating the state $|\Psi_{b,\tau_0}\rangle$. This state (evolved with the usual Hamiltonian for the 4D theory) is dual to a Lorentzian geometry that is the analytic continuation of the Euclidean wormhole described above (see Figure \ref{fig:cosmo}b,c,d).\footnote{In this context, the Euclidean wormhole is interpreted as a ``bra-ket wormhole''.} This will generally be a flat FRW big-bang/big-crunch cosmology. In the case where the 4D field theory is holographic, the 4D cosmological physics is confined to an ETW brane living at the IR end of a five-dimensional asymptotically AdS spacetime. This lies behind the horizon of a (planar) black hole, emerging from the past singularity and ending up in the future singularity (Figure \ref{fig:cosmo}d). But when the 4D theory is not a conventional holographic theory, there is no geometrical 5D spacetime. In the language of \cite{Almheiri:2019hni,Almheiri:2019yqk,Hartman:2020khs}, we can think of the 4D cosmological spacetime is an ``island'' whose physics is encoded in the state of a field theory that is not conventionally holographic.\footnote{This is similar to the ideas in \cite{Simidzija:2020ukv}, which argued that bubbles of a spacetime associated with some holographic CFT can be encoded in states of a different CFT, which might have a significantly smaller central charge.}

\subsubsection*{Eternally traversable wormholes}

\begin{figure}
\centering
\includegraphics[width=40mm]{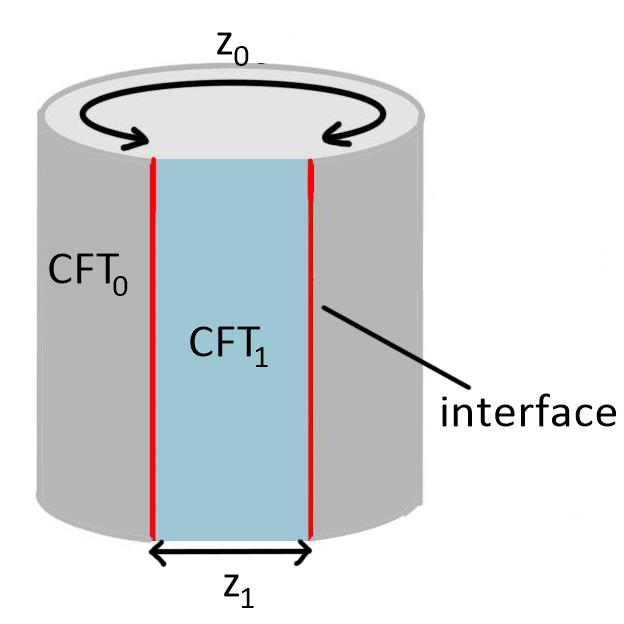}
\caption{CFTs on $R^{2,1}$ times $S^1$, with each CFT covering an interval on the $S^1$. A holographic model suggests that the negative Casimir energy of the CFT with larger central charge can become much larger than that for this CFT on $R^{2,1} \times S^1$ for special choices of the interface corresponding to a bulk interface tension close to a lower critical value in the holographic model. }
\label{fig:negen}
\end{figure}

If they exist, the Euclidean wormholes we describe (for the case of spatial $\mathbb{R}^3$) could instead be analytically continued along one direction of the $\mathbb{R}^3$ to give a four-dimensional eternally traversable wormhole (Figure \ref{fig:dual}d, but with one of the translationally invariant directions analytically continued to give a time direction). The existence of such solutions in the effective description requires a substantial violation of the averaged null energy condition \cite{Galloway:1999bp}. In the 1+1-dimensional construction of Maldacena and Qi \cite{Maldacena:2018lmt}, this arises through a direct coupling of the CFTs associated with the asymptotic regions. It has been argued in \cite{Freivogel:2019lej} and \cite{LinMaldacena} that obtaining the required amount of negative null energy in a higher-dimensional construction is difficult. We recall these arguments in detail in Section \ref{sec:effective}, and explain a possible mechanism to produce the required negative energy in the effective description. We argue that the matter in the effective description can be modeled as an interface theory as shown in Figure \ref{fig:negen}; a holographic model suggests that such setups can lead to large negative Casimir energy densities for interfaces with specific properties. This will be discussed in more detail in \cite{MSV}.

\subsubsection*{Microscopic construction}

\begin{figure}
\centering
\includegraphics[width=150mm]{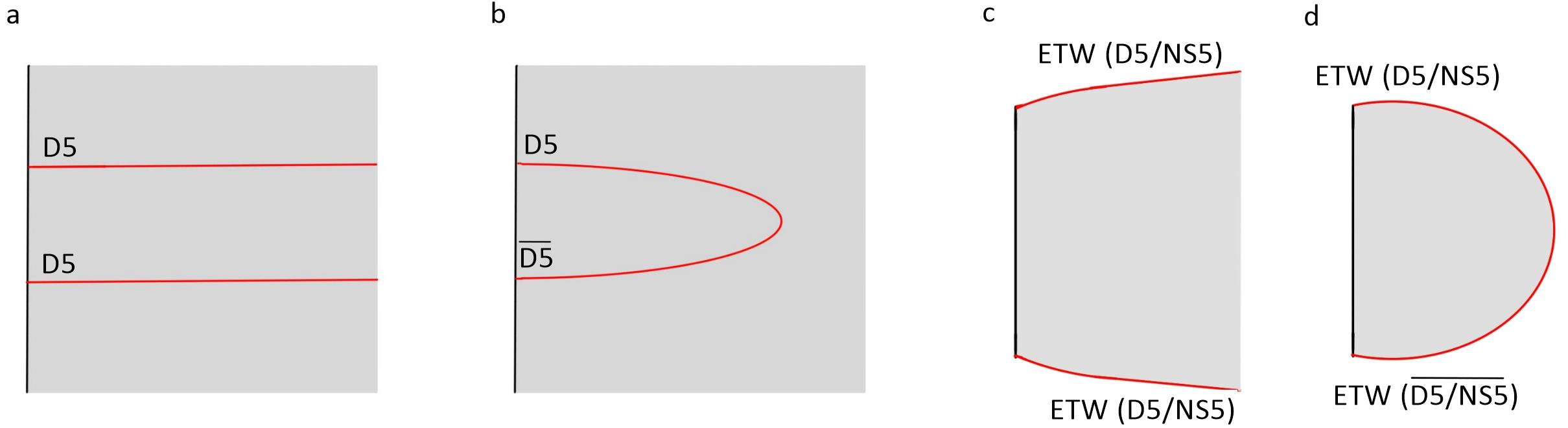}
\caption{(a) Probe brane solution dual to ${\cal N}=4$ SYM with parallel D5-brane defects. (b)Probe brane configuration for parallel defects with opposite orientation (D5-$\bar{D5}$). (c) Dual gravity solution for SUSY-preserving BCFT, with ETW branes that stay separated. (d) Suggested dual gravity solution with boundary SCFTs of opposite orientation, breaking SUSY. }
\label{fig:ddbar}
\end{figure}

The generalities of our construction are motivated and described more fully in section 2 below. In order to make everything more concrete, we discuss a possible specific realization of the construction within string theory in section 3. Here, the field theory arises as the low-energy limit of D3-branes stretched between a D5-brane/NS5-brane stack and a complementary $\overline{\rm D5}$-brane/$\overline{\rm NS5}$-brane stack, with extra D3-brane degrees of freedom added to the fivebrane stacks at either end (see Figure \ref{fig:newquiver}). The low-energy field theory description is $U(N)$ ${\cal N}=4$ SYM theory on $\mathbb{R}^3$ times an interval, coupled to 3D superconformal field theories at either end of the interval. These SCFTs are holographic with many more local degrees of freedom than the  ${\cal N}=4$ theory. These SCFTs individually preserve half the supersymmetry when coupled to the ${\cal N}=4$ theory. However, the full construction breaks supersymmetry. We argue that the theory still has a gravitational dual well-described by type IIB supergravity, and we describe the asymptotic geometry explicitly using the work \cite{DHoker:2007zhm,DHoker:2007hhe,Aharony:2011yc,Assel:2011xz}. In these geometries, the ETW branes are geometrical, characterized by an internal space which grows in size before pinching off smoothly (Figure \ref{fig:bagpipe}). We can think of them as a geometrized stack of branes emerging from one boundary and a geometrized stack of the corresponding anti-branes emerging from the other boundary. The ETW branes connecting up would then be a non-perturbative geometrized version of the joining of probe D5 and anti-D5 branes associated with parallel defects in the ${\cal N}=4$ theory (Figure \ref{fig:ddbar}) \cite{Antonyan:2006pg,Davis:2011am,Grignani:2014vaa}.

\section{General construction}

In this section, we describe and motivate the general construction in more detail before turning to the specific microscopic construction in section 3.

\subsection{Euclidean wormholes, eternal traversable wormholes, and cosmology}

Our goal is to construct models of big bang cosmology using the tools of AdS/CFT. Maldacena and Maoz pointed out that certain big-bang / big-crunch spacetimes arise by analytic continuation from Euclidean AdS wormholes, with geometry of the form
\be
\label{EucW}
ds^2 = d \tau^2 + f(\tau) ds_M^2
\ee
where $M$ is a homogeneous isotropic space and the geometry is asymptotically AdS${}^4$ for $\tau \to \pm \infty$. The form of this Euclidean geometry suggests that it could be related holographically to a pair of Euclidean CFTs, each living on $M$. However, for a pair of decoupled CFTs, the partition function and all correlation functions would factorize between the two CFTs, while holographic calculations in the geometry (\ref{EucW}) would give non-factorizing results.

There is another puzzle with the geometries in (\ref{EucW}) that applies to the flat case. Here, we could analytically continue one of the spatial directions in $M = \mathbb{R}^3$ to obtain a static Lorentzian geometry with two asymptotically AdS regions. Such a planar traversable wormhole geometry cannot exist without violating the averaged null-energy condition (ANEC) \cite{Galloway:1999bp}.

\subsection{Coupling auxiliary degrees of freedom}

To resolve these puzzles, it has been suggested that Euclidean AdS wormholes may correspond to ensemble-averaged products of CFT partition functions (see e.g. \cite{Maldacena:2004rf, Saad:2019lba, Marolf:2020xie, Maloney:2020nni, Afkhami-Jeddi:2020ezh, Cotler:2020ugk}), or partition functions for CFTs that are weakly interacting in some way (see e.g. \cite{Betzios:2019rds}). Either of these can explain the non-factorization of correlators, and for the flat case, it is understood that introducing interactions between the CFTs associated with asymptotic regions can give ANEC-violating matter in the bulk that allows a traversable wormhole\cite{Gao2016,Maldacena:2017axo,Maldacena:2018lmt}.\footnote{Note, however that there has not been an explicit construction of an eternal traversable wormhole in four dimensions with $\mathbb{R}^3$ spatial slices.}

A specific approach that incorporates features of ensemble averages and interactions is to couple the original CFTs to some auxiliary degrees of freedom spread over an extra spatial dimension \cite{VanRaamsdonk:2020tlr}.\footnote{See \cite{Kirklin:2020qtv} for a related construction involving the coupling of two theories via an auxiliary system.} Specifically, we can consider a four-dimensional CFT on $M$ times a spatial interval $[-\tau_0/2,\tau_0/2]$, with the fields at $\tau = \pm \tau_0$ coupled to our original CFTs (Figure \ref{fig:basic}). The partition function for the full theory can be understood as a product of partition functions of the original CFTs, averaged over an ensemble of sources \cite{VanRaamsdonk:2020tlr}. Here, the sources are fields in the auxiliary theory and the probability distribution for the sources comes from the path integral over the auxiliary degrees of freedom.

In order that the dual gravitational system associated with the coupled CFTs remains effectively four-dimensional, we require that the number of local degrees of freedom in the auxiliary CFT is small compared to the number of local degrees of freedom in the original CFTs. In this case, the addition of auxiliary degrees of freedom can be understood to be a small perturbation of the original theory, at least in the UV.\footnote{Below, it will be important that such perturbations can significantly alter the IR physics.}

\subsubsection*{Localized gravity and Karch-Randall branes}

To understand the effects on the gravitational physics from coupling to a 4D auxiliary CFT, it is helpful to consider the case where these auxiliary degrees of freedom are also holographic. Consider first the case where we have a single 3D holographic CFT which we couple to a four-dimensional CFT on a half-space. In this case, the full dual geometry has an asymptotically AdS${}^5$ region, and the original four-dimensional gravitational theory describes the physics of an end-of-the-world brane, as shown in Figure \ref{fig:dual}b.\footnote{In microscopic examples, the full geometry can be understood as a warped product of AdS${}^4$ and an internal space. When we have only the 3D CFT, the internal space is compact. Coupling to the 4D CFT on a half space modifies this compact space to include a narrow semi-infinite throat.\footnote{This structure was described as a ``bagpipe'' in \cite{Bachas:2018zmb}}.} Gravity is localized to this end-of-the-world brane via the Karch-Randall mechanism \cite{Karch:2000ct}. Specifically, the physics of the ETW brane has an effective description as four-dimensional gravity, where the 4D graviton obtains a tiny mass ($m^2 \sim c_4/c_3$), and we have a tower of massive fields coming from the modes of the 5D-graviton \cite{Porrati:2001gx,Duff:2004wh}. In a bottom-up description, we can think of the brane as a hypersurface living at some angle $\theta$ in the Poincar\'e coordinates of AdS${}^5 \times$S${}^5$. When this angle is close to $-\pi/2$ as in Figure \ref{fig:dual}b, the brane acts as a cutoff surface in AdS, and the effective description of the physics on the brane will include a cutoff version of the 4D CFT.

When we have two 3D holographic CFTs coupled by a 4D holographic CFT, the dual geometry now includes two ETW branes in the UV. We will argue below that in some cases, these connect up in the IR (Figure \ref{fig:dual}d), giving rise to a wormhole in the effective description.

\subsubsection*{Requirement for strong IR correlations}

We have emphasized that the number of auxiliary degrees of freedom should be small in order to maintain the four-dimensional character of the dual gravitational theory. However, ending up with a connected wormhole means that the interactions between the two CFTs induced by the auxilary degrees of freedom lead to large correlations. In particular, we require that $\ln(Z/(Z_1 Z_2)) \sim c_{3D}$. In the Lorentzian case where we have analytically continued one of the directions of $M = \mathbb{R}^3$, the geometrical connection between the two sides implies that the vacuum entanglement between the original CFTs induced by the auxiliary degrees of freedom is large, with entanglement entropy of order $c_{3D}$.\footnote{To regulate the entanglement entropy, we can consider a subsystem including a ball-shaped region of one of the 3D theories and compare the entanglement entropy of this region to the entanglement entropy for the same region in the case where there is not a second 3D CFT.}

Thus, we wish to introduce an auxiliary theory whose number of degrees of freedom is small, $c_{4D} \ll c_{3D}$, but which leads to entanglement/correlations between the original CFTs that are large, of order $c_{3D}$. In the next section, we will argue that the physics of RG flows and confinement may help achieve this.

\subsection{Confinement and symmetry breaking}

Our general Euclidean field theory setup has two 3D CFTs on a homogeneous and isotropic space $M$ coupled to a 4D CFT on $M$ times an interval $I_{\tau_0} = [-\tau_0/2,\tau_0/2]$. We would like to understand whether for some $\tau_0$, our field theory has a dual gravitational description as a 4D Euclidean AdS wormhole. We expect that the correlations between the original CFTs will become larger for smaller values of $\tau_0$, so it is natural to consider the small $\tau_0$ limit and ask whether the wormhole exists here. Since the field theories we are dealing with are assumed to be scale-invariant, we can equivalently keep $\tau_0$ fixed and take the curvature length scale of $M$ large, so that our field theory geometry approaches $\mathbb{R}^3 \times I$. If the wormhole exists in this flat case, it should also exist in the spherical and hyperbolic cases for sufficiently small spatial curvature.

At length scales much larger than $\tau_0$, the Euclidean field theory on $\mathbb{R}^3 \times I_{\tau_0}$ or the related Lorentzian field theory on $\mathbb{R}^{2,1} \times I_{\tau_0}$ should be described by some three-dimensional field theory. The IR limit of this theory could either be a non-trivial 3D CFT, or it could be a gapped theory. We will now argue that the latter may correspond to some type of connected ETW brane wormhole geometry in the dual gravitational description.

To see this, it is helpful again to consider the case where the auxiliary degrees of freedom are holographic. In this case, we have a dual geometry with an asymptotically AdS${}^5 \times$S${}^5$ region whose boundary geometry is $\mathbb{R}^3 \times I_{\tau_0}$, and we have ETW branes in the bulk anchored to the ends of the interval.

In the case where the IR theory is conformal, we have non-trivial physics at arbitrarily long wavelengths, and the bulk picture is that the radial direction extends to infinite distance in the IR. Here, the ETW branes can remain separate and extend infinitely into the IR (Figure \ref{fig:dual}c). Alternatively, they could join up somehow and extend into the IR. In neither case do we get the desired wormhole geometry.

\begin{figure}
\centering
\includegraphics[width=80mm]{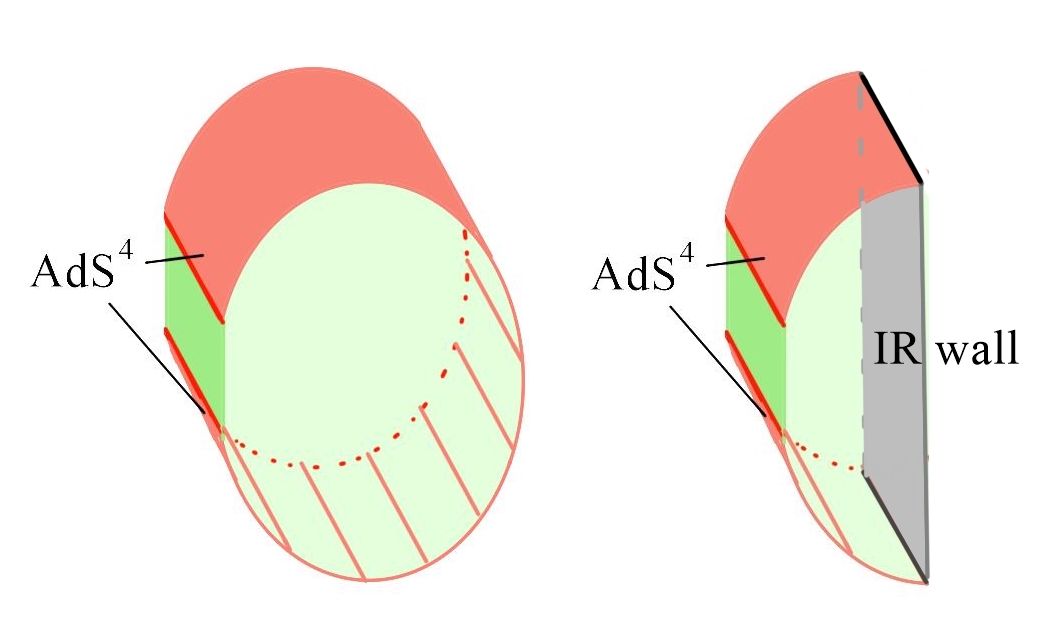}
\caption{Two possibilities for the gravity dual of a theory that confines in the IR. These can sometimes be distinguished by a different pattern of global symmetry breaking. The ETW brane geometry in the left case is a Euclidean wormhole.}
\label{fig:confining2}
\end{figure}

On the other hand, when the field theory is gapped in the IR, we expect that the radial direction should terminate somehow in the IR at a finite distance from any interior point of the geometry. A natural way for this to occur is for the two ETW branes to join up into a single brane (Figure \ref{fig:confining2}, left). In this case, the ETW brane worldvolume geometry is the desired asymptotically AdS Euclidean wormhole.

The configuration of Figure \ref{fig:confining2} (left) is not the only way to end up with a confining theory. Indeed, adding a relevant deformation to the individual 3D theories, or to the 4D CFT could lead to confinement. In this case, the ETW brane geometries and the bulk geometry could individually truncate in the IR, as shown in Figure \ref{fig:confining2}, right. Here, the effective 4D description of the ETW brane physics on the gravity side would have two disconnected asymptotically AdS spacetimes with an IR end.

In order to ensure a single connected ETW brane, one strategy is to endow the ETW brane with properties similar to those of a string theory brane, such that the pair of ETW branes acts like a probe brane-antibrane pair. For example, we can take the 3D CFTs to each include some global symmetry $G$ and the theories to be related to each other by reflection through $\tau = 0$. The $G \times G$ global symmetry in the UV is related to the presence of bulk gauge fields associated with the ETW branes. If the branes connect up in the bulk the $G \times G$ global symmetry is broken to a single diagonal copy. We will provide explicit examples below.

\subsection{From two-sided black hole to eternally traversable wormhole}

\begin{figure}
\centering
\includegraphics[width=140mm]{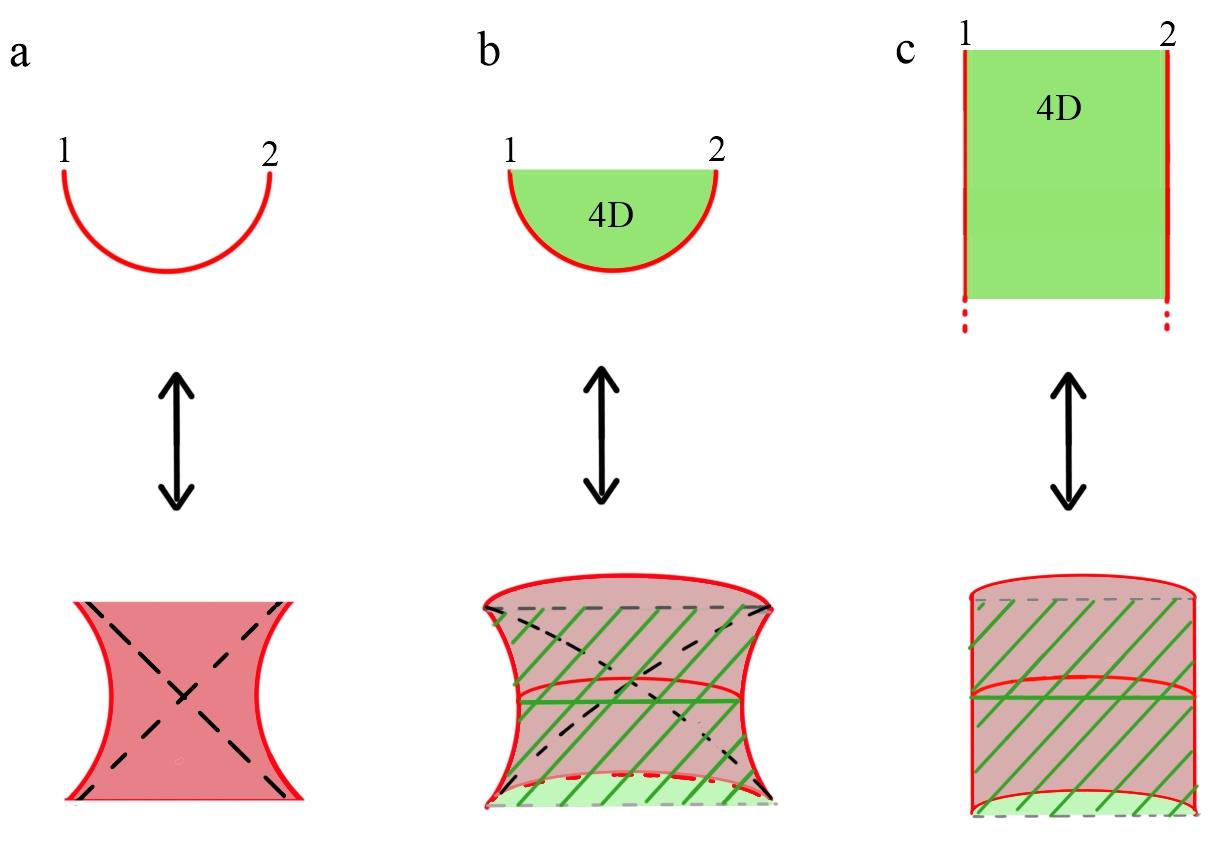}
\caption{States created by Euclidean path integrals and gravity duals. (a) Thermofield double state of two 3D CFTs, dual to a two-sided black hole. (b) State of a pair of 3D CFTs coupled by a 4D CFT. The two-sided black hole is now the geometry of an ETW brane. The black hole may be traversable for some time. (c) Vacuum state of the 3D CFTs coupled by a 4D CFT, if the ETW brane remains connected and gives an eternally traversable wormhole in the effective description.}
\label{fig:tfdtoetw}
\end{figure}

Before turning to specific microscopic models, we motivate the existence of a connected wormhole in a different way. Here, we focus on the Lorentzian picture, where we would have an eternally traversable wormhole after analytically continuing on one of the directions of the $\mathbb{R}^3$.

Consider first the pair of 3D CFTs on spatial $\mathbb{R}^2$ in the thermofield double state. This is dual to the single connected geometry of a two-sided planar 4D black hole. This state may be constructed using a Euclidean path integral with path integral geometry $\mathbb{R}^2 \times I$ that connects the two spatial $\mathbb{R}^2$s, as shown in Figure \ref{fig:tfdtoetw}a.

Next, we can consider coupling the 3D CFTs via a 4D CFT as above. For the coupled theory, we can consider the path-integral state shown in Figure \ref{fig:tfdtoetw}b. We can choose to evolve this state forward using the time independent Hamiltonian for the 4D theory on spatial $\mathbb{R}^2 \times I$. When the 4D CFT has many fewer degrees of freedom than the 3D CFTs, we expect that the gravitational description of the new state is in some sense a small perturbation of the original two-sided black hole geometry. In particular, we expect that the ETW brane geometry for the $t=0$ spatial slice should be almost the same as that of the two-sided 4D planar black hole and the ETW branes from the two different 3D CFTs should still connect. When the 4D theory is holographic, we can visualize the full geometry as having a 5D bulk such that the original 4D black hole becomes an ETW brane in this geometry. The presence of the 4D CFT may alter the time-dependence of the ETW brane. Since it represents a coupling between the original 3D theories, it can have the effect of making the ETW brane geometry traversable. However, the state is still time-dependent.

By continuously modifying the path integral geometry to the strip geometry of figure Figure \ref{fig:tfdtoetw}c, we end up with the vacuum state of the theory with the two 3D CFTs coupled by the 4D auxiliary theory. In the case we are interested in, the ETW brane geometry would remain connected in the limit where we reach the vacuum state. Since the final dual geometry is static, the ETW brane geometry should be an eternally traversable wormhole, and after analytic continuations give a Euclidean AdS wormhole and a flat cosmological spacetime.

Of course, for some theories, it could be that the ETW brane disconnects in the limit where the state approaches the vacuum state. Our goal is to find examples where the ETW brane remains connected in this limit.

\section{Microsopic construction}

In this section, we describe a family of specific microscopic constructions designed to realize the picture we have described. In order to have the largest amount of control, we take as building blocks quantum field theories with large amounts of supersymmetry, though this supersymmetry will end up being broken in the final construction.

For simplicity and maximal control over the gravity picture, we start by choosing the $U(N)$ ${\cal N}=4$ SYM theory as the 4D CFT that gives our auxiliary degrees of freedom. Here, $N^2 (= c_{4D})$ and the 't Hooft coupling $\lambda$ can both be taken large if we wish to have a 5D gravity dual, but we can also consider the case where they are not large.

\begin{figure}
\centering
\includegraphics[width=150mm]{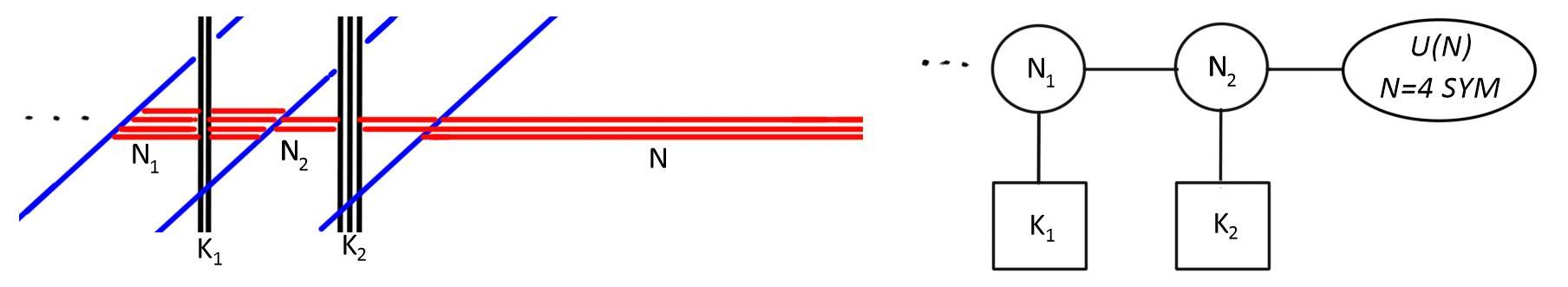}
\caption{Left: Brane construction for ${\cal N}=4$ SYM coupled to a 3D superconformal gauge theory. D3-branes in the 0123 directions (horizontal, red) are streched between D5-branes in the 0456 directions (vertical, black) and NS5-branes in the 0789 directions (blue, angled). Right: quiver gauge theory describing the low-energy physics. Circles represent circles represent gauge theory sectors coupled by bifundamental hypermultiplets (horizontal lines). Squares represent additional fundamental hypermultiplets.}
\label{fig:newquiver}
\end{figure}

Next, we introduce the 3D holographic CFTs. We take these to be superconformal theories that can be coupled to the ${\cal N}=4$ theory at a boundary while preserving half of the original supersymmetry of the ${\cal N}=4$ theory. Such theories preserve $OSp(2, 2|4)$ superconformal symmetry; they were classified by Gaiotto and Witten in \cite{Gaiotto:2008ak} (see also \cite{Nishioka:2011dq}). The BCFT obtained by coupling one of these theories to the $U(N)$ ${\cal N}=4$ theory describes the low-energy physics of $N$ semi-infinite D3-branes (in the 0123 directions) ending on stacks of D5-branes (in the 123456 directions) and NS5-branes (in the 123789 directions), with additional D3-branes stretched in the 3 direction between the D5s and NS5s, as shown in Figure \ref{fig:newquiver}. The 3D theory on its own corresponds to the physics of these extra D3-branes. Since we are free to add an arbitrarily large number of these, we can take $c_{3D}/c_{4D}$ as large as we want.

The 3D theories can also be understood as the IR limit of certain supersymmetric quiver gauge theories of the type shown in Figure \ref{fig:newquiver} (right). The parameters describing the quiver -- the ranks of the gauge groups and the number of fundamental hypermultiplets -- are related to the numbers of D5-branes, NS5-branes, and D3-branes in the string theory construction.

The dual gravity solutions for both the 3D SCFTs and the BCFTs obtained by coupling these to the ${\cal N}=4$ theory are known explicitly. These solutions of type IIB supergravity were described in \cite{DHoker:2007zhm, DHoker:2007hhe, Aharony:2011yc, Assel:2011xz}, based on the general $OSp(2, 2|4)$-symmetric solutions of type IIB supergravity found in \cite{DHoker:2007zhm, DHoker:2007hhe}.

For our construction, we want ${\cal N} = 4$ SYM theory to couple to a 3D CFT at each end of the Euclidean time interval. From the string theory perspective, we can obtain such a theory by introducing additional stacks of D5-branes and NS5-branes so that the D3-brane stack has two boundaries. The distance between the boundaries can be scaled so that we end up with a finite separation between the two boundaries in the low-energy limit. We can preserve supersymmetry if the new D5-branes and NS5-branes have the same orientation as the original ones. However, we instead want to take them to have the opposite orientation, i.e. to use anti-branes instead of branes. From the field theory perspective, what we want is to take the same Euclidean SCFT at either end of the Euclidean time interval, but coupled with the opposite orientation to the ${\cal N}=4$ theory. The reason is that we want a construction that is symmetric under Euclidean time reversal. This is required for our interpretation of the Euclidean geometry as a bra-ket wormhole, and ensures that we obtain a real Lorentzian geometry under analytic continuation. Choosing the boundary theories to have the opposite orientation breaks supersymmetry in the resulting low-energy field theory, but only nonlocally, due to boundary conditions in the ${\cal N} = 4$ theory that are mutually incompatible with supersymmetry.

We will argue that the resulting non-supersymmetric theory should be gapped in the IR, and that the dual gravity interpretation has a connected ETW brane whose effective description can be a four-dimensional Euclidean AdS wormhole.

\subsection{Dual geometries for the single-boundary case}

Before discussing the complete two-boundary construction, let us describe more explicitly the dual gravitational physics of the single-boundary theories that preserve supersymmetry, following \cite{DHoker:2007zhm, DHoker:2007hhe, Aharony:2011yc,VanRaamsdonk:2020djx}.

\begin{figure}
\centering
\includegraphics[width=120mm]{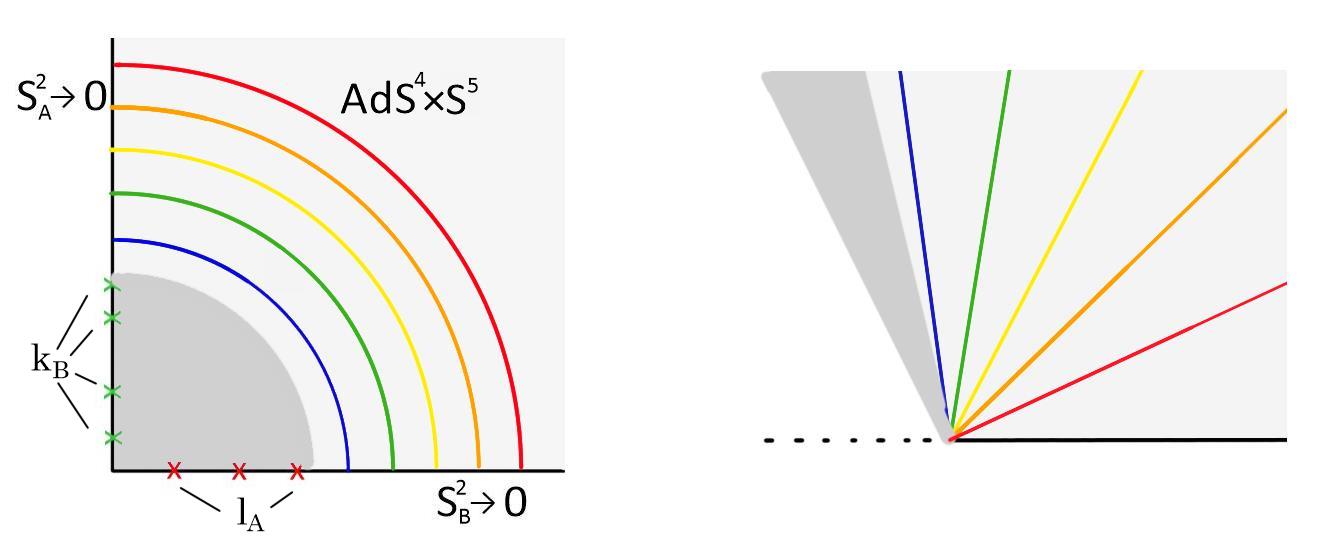}
\caption{Left: Geometries dual to ${\cal N} = 4$ on a half-space. Each point in the quadrant has an $AdS^4 \times S^2 \times S^2$ fiber. Curves connecting the axes are topologically $AdS^4 \times S^5$. Poles on the $x$ and $y$ axis correspond to D5 and NS5-brane throats. Right: Full geometry is well-approximated by Poincar\'e-AdS away from the dark grey shaded region, where the internal space smoothly degenerates. This region can be understood as an end-of-the-world brane from the lower-dimensional perspective.}
\label{fig:asymptotic}
\end{figure}

The bosonic symmetry of the full BCFT includes the $SO(3,2)$ 3D conformal symmetry plus an $SO(3) \times SO(3)$ subset of the original $SO(6)$ R symmetry. Accordingly, the dual geometries takes the form of $AdS^4 \times S^2 \times S^2$ fibered over a two-dimensional space. In general, we can write the metric as
\be
f_4(r,\theta) ds^2_{AdS_4} + f_1(r,\theta) d \Omega_2^2 + f_2(r,\theta) d \Omega_2^2 + 4 \rho(r,\theta)(dr^2 + r^2 d \theta^2)
\ee
where $r$ and $\theta$ are polar coordinates on the first quadrant of a plane. This is illustrated in Figure \ref{fig:asymptotic}. The metric functions $f_4,f_1,f_2$, and $\rho$ are determined by a pair of harmonic functions $h_1,(r,\theta),h_2(r,\theta)$, and these are determined by choosing the locations $\{l_A\}$ for a set of poles of $h_1$ on the $x$ axis and the locations $\{k_B\}$ for a set of poles of $h_2$ on the $y$ axis, where multiplicities are allowed. The explicit form of the metric for these solutions is reviewed in Appendix \ref{app:microscopic}; see the references \cite{DHoker:2007zhm, DHoker:2007hhe, Aharony:2011yc,VanRaamsdonk:2020djx} for more details, including the expressions for the other supergravity fields.

The geometry is illustrated in Figure \ref{fig:asymptotic}. At each point in the quadrant, we have an $AdS^4 \times S^2 \times S^2$ fiber, where the volumes of the three factors can vary independently. The first and second $S^2$ volumes go to zero for $\theta = 0$ and $\theta = \pi/2$ respectively, except at the locations of the poles, which are associated with $D5$-brane throats ($x$-axis poles) or $NS5$-brane throats ($y$-axis poles) in the geometry. The pair of $S^2$s fibred over a curve connecting the two axes (e.g. a constant $r$ curve) gives a geometry that is topologically $S^5$. For large $r$, the curves of constant $r$ describe $AdS^4 \times S^5$ slices of the local $AdS^5 \times S^5$ geometry that describes the asymptotic region. These are the slices of fixed Poincar\'e angle, as shown on the right in Figure \ref{fig:asymptotic}.

In the region where the solution is well-described by Poincar\'e $AdS^5 \times S^5$, the variable $r$ is related to the angular coordinate $\Theta_p$ in the $\tau-z$ plane in Poincar\'e coordinates via
\be
\label{Poinc}
{ r \over r_0} = {1- \sin \Theta_p \over \cos \Theta_p} \qquad \qquad ({\rm large} \; r) ;\ .
\ee
For smaller values of $r$, the geometry deviates from $AdS^5 \times S^5$; the $r=0$ point corresponds to a smooth part of the geometry where the $S^5$ contracts to zero size.

The pole locations are constrained in the microscopic type IIB string theory by the requirement that the various fluxes originating from the fivebrane throats should be quantized. Specifically, we must have that
\bea
\label{fluxconstr}
L_A &=&  \sqrt{g} l_A + {2 \over \pi} \sum_B \arctan{l_A \over k_B} \cr
K_B &=&  {k_B \over \sqrt{g}} + {2 \over \pi} \sum_A \arctan{k_B \over l_A}
\eea
are integers for each $A$ and $B$. These integers are related to the number of units of D3-brane flux per fivebrane in a given throat, and are directly related to integer parameters in the brane or quiver pictures of Figure \ref{fig:newquiver} specifying the gauge theory.

\subsubsection*{Microscopic picture of the ETW brane}

The geometry for a given microscopic 3D SCFT labeled by parameters $\{L_A\},\{K_B\}$ and corresponding supergravity parameters $\{l_A\},\{k_B\}$ will contain a portion that is a good approximation to the part of $AdS^5 \times S^5$ with Poincar\'e angle  $\Theta_p > \Theta_0$ for some angle $\Theta_0$ that is different for different parameter choices.\footnote{Here, $\Theta_0$ that depends on how closely we require the geometry to match with $AdS^5 \times S^5$}. The remainder of the geometry (grey shaded region in Figure \ref{fig:asymptotic}) can be understood as a fat ETW brane in which the $S^5$ contracts. This part of the geometry also includes the fivebrane throats. We can think of $\Theta_0$ as the ETW brane angle.

We will now argue that by choosing the boundary SCFT appropriately, we can find examples with $c_{3D} \gg c_{4D}$ where $\Theta_0$ is arbitrarily close to $-\pi/2$, so that our geometry includes an arbitrarily large portion of $AdS^5 \times S^5$. In this case, the ETW brane is like a Planck brane cutting off the asymptotic region on half the space, and we expect that gravity should be well localized on the brane.

First, we note that the number and location of the poles determines the rank $N$ of the ${\cal N}=4$ SYM theory gauge group, and the asymptotic AdS radius $L$ by
\be
N = {L^4 \over 4 \pi \ell_s^4} = \sum_A l_A + \sum_B k_B \; .
\ee
For the solution specified by parameters $\{l_A\}$ and $\{k_B\}$, we can expand the metric functions asymptotically in $r$ to verify that the solution asymptotes to $AdS^5 \times S^5$ with these parameter values. The same asymptotic behavior is obtained for many different choices of poles; these choices correspond to our choice of 3D SCFT.

For fixed $N$, we find that the ETW brane angle $\Theta_0$ can be made to approach $-\pi/2$ by taking the pole locations $\{l_A\}$ and $\{k_B\}$ small compared with
\be
r_0 = \sqrt{N} = \sqrt{\sum_A l_A + \sum_B k_B} \; .
\ee
This requires taking a large number of poles. Since each pole corresponds to a fivebrane in the brane construction, these cases correspond to having a complicated 3D SCFT with $c_{3D} \gg c_{4D}$.\footnote{As a specific example (setting $g=1$), we can take a pole with multiplicity $N_5$ at location $k = N/(2N_5)$ and a pole with multiplicity $N_5$ at location $l = N/(2 N_5)$. The flux quantization constraints (\ref{fluxconstr}) require that $N/(2N_5) + N_5/2$ is an integer.  We find that for $N_5 \gg N^{1 \over 2}$, the solution includes a region that is a good approximation to the portion of $AdS^5 \times S^5$ with $\theta < \pi/2 - \epsilon$, where $\epsilon = N^{1 \over 4}/ \sqrt{N_5}$. Note that since $N_5$ can be as large as $N$, we can make $\epsilon$ parametrically small.}

Thus, we find that by choosing a boundary SCFT with many degrees of freedom, the effective ETW brane tilts strongly outward so that it should behave like a Planck brane cutting off half of the asyptotic region of Poincar\'e-AdS. We expect gravity to localize on the brane via the Karch-Randall mechanism. The resulting theory is expected to have an effective description as the ${\cal N}=4$ theory on a half-space coupled to a theory on the other half-space that includes the gravitational theory dual to our 3D SCFT and a cutoff version of the ${\cal N}=4$ theory.

Even when $\Theta_0$ is not close to $-\pi/2$, the ETW brane may be effectively described by 4D gravity provided that $c_{3D} \gg c_{4D}$. As argued in \cite{Bachas:2018zmb}, this generally corresponds to a situation where the internal space volume in the ETW region becomes large before contracting; such a geometry (shown in Figure \ref{fig:bagpipe} (left)) was described in \cite{Bachas:2018zmb} as a ``bagpipe''. Here, the ``bag'' without the pipe is the internal space geometry for the dual of the 3D SCFT without the ${\cal N}=4$ theory. Coupling to the ${\cal N}=4$ theory adds the pipe, and this is very narrow compared to the bag when $c_{3D} \gg c_{4D}$. From the effective field theory perspective, the addition of the pipe gives the 4D graviton a small mass $m^2 \sim c_{4D}/c_{3D}$, and adds a tower of higher-mass modes coming from the 5D-graviton. But the physics is still a small perturbation to the original 4D theory dual to our 3D SCFT.

\subsection{The two-boundary case}

Next, we consider the case with two boundaries, where we have ${\cal N}=4$ SYM theory on $\mathbb{R}^3 \times I$ with 3D superconformal field theories of opposite orientation on either side of the interval. In this case, each SCFT preserves a different half of the supersymmetries of the ${\cal N}=4$ theory, so the full theory has no remaining supersymmetry. Before discussing this case, it will be useful to understand also the case where the two SCFTs preserve the same supersymmetries.

\subsubsection*{Aside: wedge holography and a 3D dual for $AdS^5 \times S^5$}

The two-boundary theories preserving supersymmetry case can be understood as arising from a string theory construction with D3-branes stretched between two separate stacks of D5-branes and NS5-branes, where we adjust the length of the D3-branes to remain finite in the decoupling limit. In this case, the dual supergravity solution should include a wedge of $AdS^5 \times S^5$ with ETW branes on either side. In the IR limit, the field theory will flow to a single 3D SCFT, namely the one associated with the string theory construction above where all the fivebranes are taken to be coincident. Such SCFTs were discussed in \cite{Bachas:2017rch}, and provide examples of the ``wedge holography'' discussed in \cite{Akal:2020wfl}, where a 3D CFT is dual to a wedge of a 5D AdS space.\footnote{Note that because of the AdS geometry, the proper distance between the ETW branes actually remains constant as a function of the radial coordinate, so we can think of this as an example of ordinary holography where the internal space includes an interval. In the microscopic examples taking into account the spheres, the full internal space takes a dumbbell shape, with two ``bags'' connected by a narrow tube \cite{Bachas:2017rch}.}

It is interesting to note that, as for the single boundary case, by a judicious choice of the boundary SCFTs, our dual geometry can include an arbitrarily large wedge of $AdS^5 \times S^5$, i.e. a wedge $-\pi/2 + \epsilon \le \Theta_p \le \pi/2 + \epsilon$ for arbitrarily small $\epsilon$. Thus, we have in a sense a 3D dual to $AdS^5 \times S^5$, though the full dual geometry also includes the ETW branes. From a field theory point of view, this suggests that the full physics of the ${\cal N}=4$ SYM theory may be contained within an appropriately chosen 3D SCFT. This may be related to the idea of dimensional deconstruction \cite{ArkaniHamed:2001ca}.

\subsubsection*{Supersymmetry breaking boundary conditions}

\begin{figure}
\centering
\includegraphics[width=120mm]{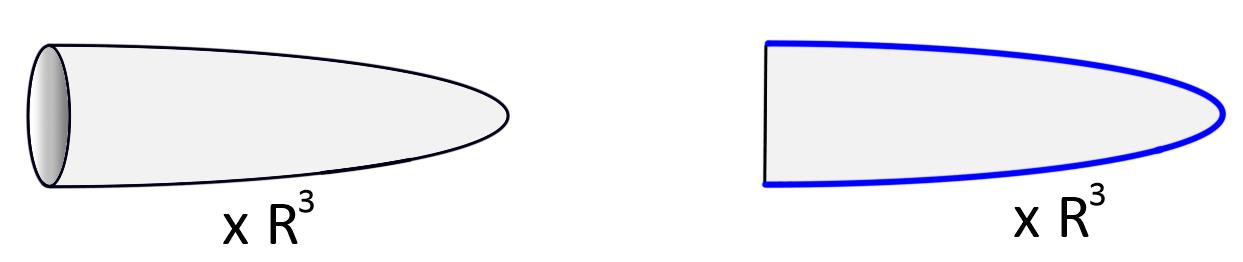}
\caption{Left: geometry dual to 4D holographic CFT on $R^3$ times a circle with antiperiodic boundary conditions for fermions (internal space is suppressed). Right: geometry dual to 4D holographic CFT on $R^3$ times an interval, with SCFT of opposite orientation at either end. The blue ETW brane is described microscopically by a smooth degeneration of the internal space.}
\label{fig:contract}
\end{figure}

We now return to the case of interest where the two boundary theories preserve non-intersecting subsets of SUSY generators, so the full theory breaks all supersymmetry (as well as some of the global symmetries present in the UV theory).

Most well-controlled microscopic examples of AdS/CFT are supersymmetric, so one may be concerned that the examples we have described cannot be studied holographically in a controlled way. However, in our case, the supersymmetry is only broken by the fact that the boundary conditions at either end of the interval are incompatible with each other from a SUSY-perspective. A simpler example where we have SUSY broken by boundary conditions is the ${\cal N} = 4$ theory compactified on a circle with antiperiodic boundary conditions for fermions, introduced by Witten. Here, the theory has a well-controlled gravity dual in which the circle becomes contractible in the bulk for the case where the noncompact directions of the field theory are $\mathbb{R}^3$.\footnote{For the case where we replace $\mathbb{R}^3$ with $S^3$, the circle is contractible in the bulk if its radius in field theory is sufficiently small compared to the $S^3$ radius. The resulting transition is the same one that appears in the Hawking-Page transition (where the circle is taken to represent Euclidean time).} From the lower-dimensional perspective, this gives us a 3D confining gauge theory \cite{Witten1998a} since the radial direction in the dual geoemtry has finite extent in the IR.

Our situation is very similar to Witten's example, except that the compact direction is an interval rather than a circle. Supersymmetry is broken by the boundary conditions, and we expect that the interval contracts and pinches off in the bulk. This can happen smoothly if the ETW branes originating from the two 3D SCFTs join up in the IR as shown in figure \ref{fig:contract}. This implies that the full geometry is capped in the IR, and the IR physics of the field theory is that of a confining/gapped 3D  theory.

It is also possible to get a gapped theory in the IR without the ETW branes connecting smoothly. The two ETW branes and the bulk geometry between then could each terminate independently in the IR (Figure \ref{fig:confining2}). Thus, we want to further motivate the idea that the ETW branes do connect in some cases.

\subsection{Probe example}

It will be useful to consider a probe example. Instead of taking D3-branes that terminate on stacks of fivebranes, we can consider D3-branes that are intersecting parallel D5-branes. In this case, the field theory description is a theory with two parallel codimension 1 defects. The physics of these defects was described in \cite{Karch:2001cw,Karch:2000gx,DeWolfe:2001pq}. We have a 3D hypermultiplet in the the fundamental representation of $U(N)$ coupled to the ${\cal N}=4$ fields at the defect \cite{DeWolfe:2001pq}. In the supergravity description, we have a probe D5-brane originating from each defect with worldvolume geometry $AdS^4 \times S^2$, where the $S^2$ lives in $S^5$.

If the defects arise from parallel D5-branes with the same orientation, the field theory preserves supersymmetry. The scale $L$ breaks conformal invariance, but we expect that the theory flows to a conformal defect theory in the infrared associated with the D3-branes intersecting two coincident D5-branes. In the dual description, we now have probe D5-branes living on parallel $AdS^4$ slices of the $AdS^5$ (Figure \ref{fig:ddbar}a).

For parallel D5-brane defects with the opposite orientation (associated with D3-branes intersecting a separated $D5-\bar{D5}$ pair), we have the same probe brane solution (replacing one $D5$ with a $\bar{D5}$) but this is now unstable, both nonperturbatively and perturbatively. To see this, recall that open strings stretched between a $D5-\bar{D5}$ pair have a mode that is tachyonic if the branes separation is smaller than the string scale. This happens in the probe solution for $z > L_{AdS} \ell/ \alpha'$, where $\ell$ is the separation in the field theory.

The endpoint of perturbative instability is another classical solution in which the branes are connected, as shown in Figure \ref{fig:ddbar}b. The explicit solutions were constructed in \cite{Antonyan:2006pg}; we review them in Appendix \ref{app:D5probe}. The physics is qualitatively similar if the probe is a small number $n$ of D5-branes and if we allow some small number $k$ of D3-branes to end on these. In this case, the probe branes tilt outward as they enter the bulk, but they still connect provided that $k/n$ is not too large. The details are presented in appendix \ref{app:boundprobe}.

\subsection{Non-perturbative version}

Starting with our probe brane setup, we can generalize to consider defects corresponding to larger numbers D5-branes or combinations of D5s and NS5s, and finally our case of interest where the D3-branes all end on the fivebranes.\footnote{In Appendix \ref{app:boundprobe}, we consider an intermediate situation with one boundary (associated with D3-branes ending on stacks of D5-branes and NS5-branes) and one defect associated with an anti-$D5$. This anti-$D5$ is a probe version of a second boundary that breaks SUSY. We find that there is a strong tendency for the probe $\bar{D5}$ in the bulk that emerges from the defect to be drawn towards the ETW brane (Figure \ref{fig:boundprobe}, right). In many case, all solutions for the probe brane are of this type. This is in contrast to the supersymmetric situation where the probe D5 stays away from the ETW brane.} In these situations, backreaction must be taken into account, and the probe brane is replaced by the geometrical ETW brane in some solution of type IIB supergravity.

\begin{figure}
\centering
\includegraphics[width=150mm]{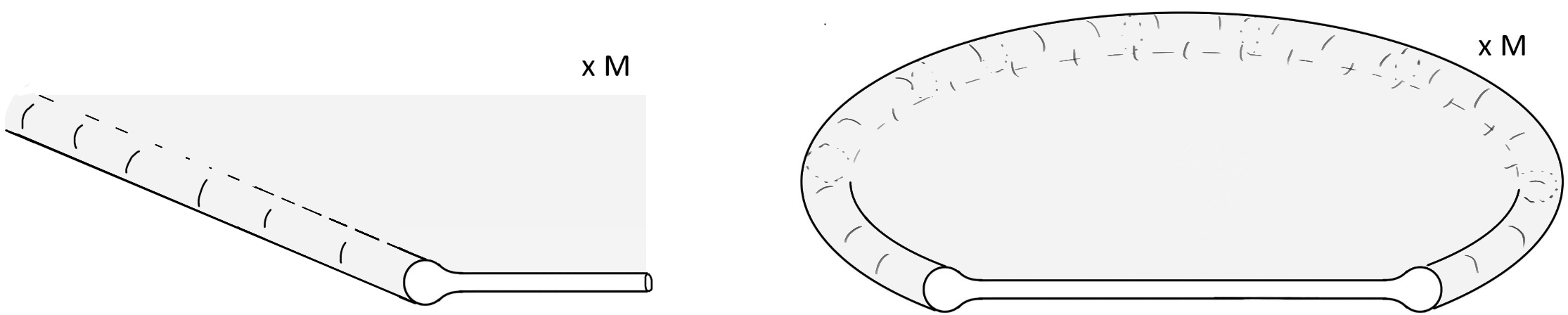}
\caption{Left: Schematic of dual geometry for ${\cal N}=4$ SYM theory on $M \times \mathbb{R}^+$ coupled to a 3D SCFT with $c_{3D} \gg c_{4D}$. Away from the ETW brane, the internal space is $S^5$. The ETW brane is a region of the 10D geometry where this is deformed, growing and then pinching off smoothly. Right: the case with ${\cal N}=4$ on an interval coupled to 3D SCFTs with opposite orientation at the ends of the interval. Shown are the Euclidean time (horizontal), radial direction (into the page) and internal space.}
\label{fig:bagpipe}
\end{figure}

It is plausible that the behavior of the ETW branes is similar to that of the probe branes. In the non-supersymmetric case where we have defects/interfaces/boundaries that are related by a reversal of orientation, we expect that the instability of the brane-antibrane configuration in the string theory picture should be reflected in a tendency for the branes to connect up in the preferred solution.  A schematic of the proposed geometry, emphasizing the geometrical nature of the ETW branes, is shown in Figure \ref{fig:bagpipe} (right).

To verify this picture, we would ideally want to look for solutions of type IIB supergravity with the appropriate asymptotic behavior, showing that a connected solution exists and that this is the solution with least action.

\subsection{Asymptotic behavior of the dual geometry}

The asymptotic behaviour of the dual geometries for our setup can be understood from the UV physics of the field theory. Correlators of bulk ${\cal N}=4$ SYM operators separated by distances much smaller than their distance to the boundary should be well-approximated by those of ${\cal N}=4$ SYM on $\mathbb{R}^4$, so the asymptotic region of the dual geometry associated with points in the field theory away from the boundaries will be $AdS^5 \times S^5$. Short-distance correlators involving operators on one of the boundaries and nearby bulk operators will be governed by the superconformal theory of ${\cal N}=4$ coupled to the 3D SCFT degrees of freedom at a single boundary. Thus, the asymptotic geometry near each of the boundaries should match with one of single-boundary solutions described above.

In the field theory with two boundaries, conformal invariance is broken, but we preserve translations and rotations in the three transverse directions. The UV theory also preserves $SO(3) \times SO(3)$ symmetry. If this is not broken spontaneously, the metric would take the form
\be
f_3(\vec{x}) d \vec{y}^2 + f_1(\vec{x}) d \Omega_2^2 + f_2(\vec{x}) d \Omega_2^2 + g_{ij}(\vec{x}) dx^i dx^j
\ee
where the three coordinates $x^i$ on which the metric functions depend correspond to the Euclidean time direction, a radial direction, and one internal direction. The other fields of type IIB supergravity will also generally be nonzero.

Since the asymptotic behavior is known, we (optimistically) expect that it is a tractable numerical problem to find the desired solutions and investigate their properties. However, this lies beyond the scope of the present investigation.

\section{Bottom-up and effective field theory descriptions}
\label{sec:effective}

An alternative to searching for the full type IIB supergravity solutions would be to look for qualitatively similar solutions in a simpler theory of gravity. The simplest possibility with Einstein gravity and a constant tension ETW brane was considered in \cite{Cooper2018}. There, it was found that connected solutions exist and have least action provided that the tension of the ETW brane is below some value $T_*$. But this value is below the critical tension $T_c$ where the Poincar\'e angle of the ETW brane approaches $-\pi/2$ and at which gravity localization takes place. For $T > T_*$, the connected ETW brane solutions are self-intersecting and don't make sense.\footnote{This is in contrast to the case of a 2D CFT, where solutions of the simple model exist for all values up to $T_c$.}

It seems likely that the non-existence of solutions for $T > T_*$ reflects a failure of the simple bottom-up model to properly capture the physics of the CFT setup. In the picture where the Euclidean path integral is preparing a state of the auxiliary degrees of freedom and we take the spatial geometry to be $S^3$, the model suggests that for all the boundary theories corresponding to $T > T_*$, the state  $e^{-\beta H} | b \rangle$ has energy of order $N^0$ even in the limit $\beta \to 0$, in conflict with the expectation that boundary states $|b \rangle$ should generally be singular. The resolution is likely that the bottom up model needs additional elements in order to properly capture the physics. The full type IIB supergravity solutions involve a non-trivial dilaton, fluxes, and an internal space that becomes larger in the vicinity of the ETW brane. There are also light degrees of freedom localized near the fivebrane throats. Likely some of these additional elements are required in a bottom-up model to properly capture the physics. Below, we will suggest a particular resolution for the problem of self-intersecting ETW branes for $T > T_*$.\footnote{A possible alternative resolution was presented by Antonini and Swingle in \cite{Antonini2019}. These authors considered adding a bulk gauge field and making the ETW brane charged under this field. In this case, connected brane solutions were found to exist all the way up to the critical tension.}

An alternative approach to understanding physics on the gravity side is to consider the effective field theory of the ETW brane. Here, the simplest possible model is to take 4D gravity coupled to a cutoff CFT (which takes the place of the 5D bulk). This should give the same physics as the bottom up model with pure 5D gravity coupled to a constant tension ETW brane. As in that description, we do not find the desired solutions in this setup \cite{Freivogel:2019lej, LinMaldacena}.\footnote{We thank Henry Lin and Juan Maldacena for emphasizing this.}  We review this analysis in the four-dimensional effective description in the next subsection. However, we expect that the correct 4D effective description should include additional elements. We note in particular that the microscopic models we consider are characterized by a global symmetry, with a symmetry breaking pattern $G \times G \to G$. In the effective field theory description, we then have a gauge field with gauge group $G$. Since the underlying theory is supersymmetric, this comes along with scalar fields and fermions. Since supersymmetry is broken by the combination of boundary conditions in our model, the vacuum energies of these fields do not cancel. Thus, it may be important to take into account the physics of these extra fields in the effective description. This will be the case in the analysis that we describe presently.

As emphasized in \cite{Freivogel:2019lej, LinMaldacena}, the main challenge in these effective models is obtaining a sufficient amount of negative energy from the matter coupled to the 4D gravity (in the picture where we are describing an eternally traversable wormhole). In the next subsection, we will review this effective field theory analysis and present a novel mechanism for achieving the large negative energy.

\subsection{Effective field theory setup}

\begin{figure}
\centering
\includegraphics[width=80mm]{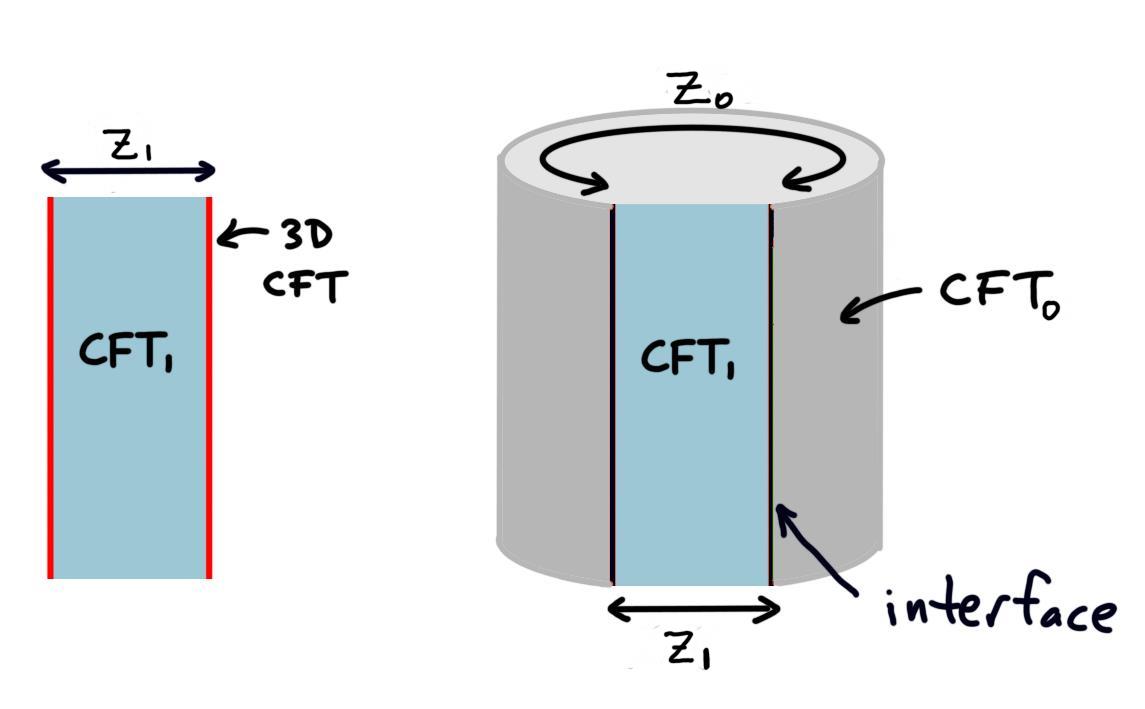}
\caption{Left: microscopic setup. Right: model for matter, in the Minkowski space conformal frame.}
\label{fig:eftsetup}
\end{figure}

We begin with our basic setup of 3D holographic conformal field theories on $\mathbb{R}^{2,1}$ coupled together by a 4D CFT on $\mathbb{R}^{2,1}$ times an interval $[-z_1/2,z_1/2]$ as in Figure \ref{fig:eftsetup} (left).

We would like to understand whether the dual description can include a connected travesable wormhole ( a connected ETW brane with localized gravity in the case where the 4D theory is holographic). We will try to come up with an effective 4D description. Our analysis is similar to that in \cite{Freivogel:2019lej, LinMaldacena}. This description should include:
\begin{itemize}
\item
The 4D gravitational theory dual to the 3D CFT, describing a spacetime with two asymptotically AdS regions. This may include additional matter.
\item
A cutoff version of the 4D CFT (accounting for the bulk physics).
\item
A non-gravitational version of the 4D CFT on a strip, which couples the fields at the two AdS boundaries.
\end{itemize}
The geometry of the ETW brane should be
\be
ds^2 = a^2(z)( dz^2 + dx^\mu dx_\mu)
\ee
where $a(z)$ has simple poles at $z = \pm z_0/2$ (giving asymptotically AdS at the two ends) and a minimum value at $z=0$. Here, the parameter $z_0$ is dynamical.

The zz-component of Einstein's equation gives
\be
3 \left({a' \over a} \right)^2 - {3 a^2\over L_{AdS}^2} = 8 \pi G T_{zz}
\ee
The stress-energy tensor comes from a cutoff version of the 4D CFT plus additional matter fields that appear in the gravity dual of the 3D CFTs. We will model the whole matter system as some 4D CFT, which we call CFT${}_0$ to distinguish it from CFT${}_1$, the original 4D CFT.

To understand the stress-energy tensor of CFT${}_0$, we can perform a conformal transformation to flat space $\mathbb{R}^{2,1} \times [-z_0/2,z_0/2]$. In this picture, the CFT${}_0$ fields are coupled at either end of the interval $[-z_0/2,z_0/2]$ to the CFT${}_1$ fields on the ends of the interval $[-z_1/2,z_1/2]$ so that the $z$ direction is periodic, as shown in Figure \ref{fig:eftsetup} (right). We can model the connection between CFT${}_1$ and CFT${}_0$ as some conformal interface.

Using the 2+1 Poincar\'e symmetry and conformal invariance, the stress tensor of CFT${}_0$ in this flat space picture must take the form
\be
T_{zz} = - {3 \over z_0^4} F\left({z_1 \over z_0} \right) \qquad T_{\mu \nu} = \eta_{\mu \nu} {1 \over z_0^4} F\left({z_1 \over z_0} \right)
\ee
where we have used the conservation and tracelessness properties, and used dimensional analysis to determine the possible dependence on $z_0$ and $z_1$, which are the only scales.

In the original conformal frame, the stress tensor becomes
\be
T_{zz} = - {1 \over a^2} {3 \over z_0^4} F \qquad T_{\mu \nu} =  \eta_{\mu \nu} {1 \over a^2} {1 \over z_0^4} F
\ee
where we are ignoring the conformal anomaly for now (we will show in Appendix \ref{app:anomaly} that it does not qualitatively change the results, though it does lead to interesting effects in the Lorentzian cosmology picture).

Then the $zz$ component of Einstein's equation gives
\be
 \left({a' \over a} \right)^2 - {a^2 \over L_{AdS}^2} = - {8 \pi G \over z_0^4 } F {1 \over a^2} \; ,
\ee
or
\be
{da \over \sqrt{{a^4 \over L^2} - {8 \pi G F \over z_0^4}}} = dz
\ee
The minimum value of $a$ occurs where $a'=0$, so we have
\be
a_{min} = {1 \over z_0} (8 \pi G F L^2)^{1 \over 4} \; .
\ee
Integrating from this minimum radius (which occurs at $z=0$) to the asymptotically AdS boundary at $z=z_0/2$, we get
\be
\label{int1}
\int_{a_{min}}^\infty {da \over \sqrt{{a^4 \over L^2} - {8 \pi G F \over z_0^4}}} = z_0/2
\ee
Defining
\be
{\cal I} = \int_1^\infty {dx \over \sqrt{x^4-1}} = {\Gamma \left({3 \over 2} \right) \Gamma \left({1 \over 4} \right) \over  \Gamma \left({3 \over 4} \right)}  = {\sqrt{2} \over 2} K\left({\sqrt{2} \over 2}\right) \approx 1.311\; ,
\ee
and rewriting the integral in (\ref{int1}) in terms of this, we get finally that
\be
\label{eqn}
F \left( {z_1 \over z_0} \right) = {2 {\cal I}^4 L^2 \over \pi G} \sim c_{3D}
\ee
This gives us an equation for $z_0$. We see that in order for solutions to exist, the function $F$ must be able to take on a large value for some $z_0$. Naively, the value of $F$ should be of order $c_0$ (the number of degrees of freedom of the matter theory coupled to gravity), which we expect to be much less than $c_{3D}$. However, in the next subsection, we will investigate the behavior of $F$ in a holographic model and show that large values $F \gg c_0$ can be achieved in certain cases.

\subsection{Holographic analysis of interface CFTs}
\label{sec:casimir}

\begin{figure}
\centering
\includegraphics[width=120mm]{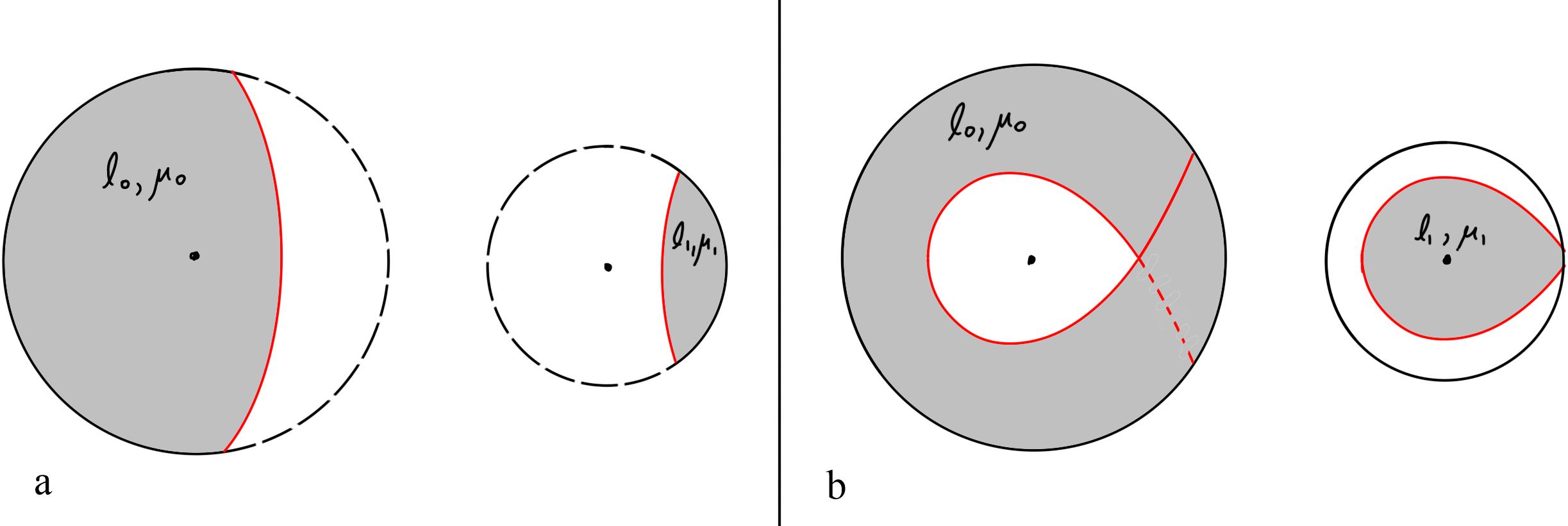}
\caption{Left: Gravity dual of the interface theory: patches of the double analytically continued AdS-Schwarzschild geometries are glued together along a constant tension domain wall. Right: when CFT${}_0$ has smaller central charge and the interface tension approaches its minimal value, the region associated with CFT${}_0$ is part of a multiple cover of the original AdS-Schwarzschild geometry.}
\label{fig:patches}
\end{figure}

In this section, we consider the CFT setup of the previous section in the case where the two CFTs are holographic. Here, we are just using holography as a tool to answer the CFT question of whether $F$ can be large. We first recall the behavior of a single holographic 4D CFT on $\mathbb{R}^{2,1} \times S^1$, where we take the $S^1$ to have length $L$ \cite{Witten1998a}. In the case where we have antiperiodic boundary conditions for fermions so that the $S^1$ is allowed to contract in the bulk, the relevant solution is a double-analytic continuation of the planar Schwarzschild geometry:
\be
\label{geom}
ds^2 = f(r) dz^2 + f^{-1}(r)dr^2 + {r^2 \over \ell^2} d x_\mu dx^\mu \; ,
\ee
where
\be
f(r) = {r^2 \over \ell^2} - {\mu \over r^2} \; .
\ee
The periodicity of the $z$ direction is fixed by smoothness at the horizon to be
\be
{\pi \ell^{3 \over 2} \over \mu^{1 \over 4}} \; ,
\ee
This should equal the CFT periodicity $L$ so we have that
\be
\mu = {\pi^4 \ell^6 \over L^4} \; .
\ee
Using the standard dictionary to read off the stress tensor, we find
\be
T_{zz} = -3 {\pi^3 \ell^2 \over 16 G} {1 \over L^4}  \qquad T_{\mu \nu} = \eta_{\mu \nu}  {\pi^3 \ell^2 \over 16 G} {1 \over L^4}
\ee
We recall that $\ell^2 / G$ gives a measure of the number of CFT degrees of freedom. In the language of the previous section, we can think of this case as having a trivial pair of interfaces with zero separation, identifying $L$ with $z_0$. In this case, we have
\be
F = {\pi^3 \ell^2 \over 16 G} \equiv c_0
\ee
so the behavior of $F$ is as expected.

Now we consider the setup of the previous section. We will employ a holographic model \cite{Simidzija:2020ukv} where the CFT interface corresponds to a constant tension domain wall between two regions with different AdS length scales $\ell_0$ and $\ell_1$ associated with CFT${}_0$ and CFT${}_1$. As explained in \cite{Simidzija:2020ukv}, the tension parameter $\kappa = 8 \pi G_5 T/3$ of the domain wall is constrained to lie between $|1/\ell_1 - 1/\ell_0|$ and $1/\ell_1 + 1/\ell_0$ in order that it can reach the AdS boundary. The parameter $\kappa$ is related to properties of the CFT interface (we can think of it as an interface central charge; this is conjectured to decrease under interface RG flows).

The dual geometries correspond to a patch of the geometry (\ref{geom}) with parameters $\ell_0, \mu_0$ connected across the domain wall to a patch of the geometry (\ref{geom}), as shown in Figure \ref{fig:patches}. The trajectory of the domain wall may be determined by solving the Israel junction conditions. The details of this analysis will be presented in \cite{SVRcasimir}, but are essentially the same as in \cite{Simidzija:2020ukv,Bachas:2021fqo}. The results for the interface trajectories are given in Appendix \ref{app:interface}. From these solutions, we can read off the behavior of the function $F$ defined in the previous section.

It will be convenient to describe the behavior of $F/c_0$ as a function of the dimensionless ratio $z_1/z_0$. For generic choices of parameters, $F/c_0$ is of order 1, with mild dependence on $z_1/z_0$. However, there is an interesting behavior when $\ell_0 > \ell_1$ and we take $\kappa$ towards the critical value $1/\ell_1 - 1/\ell_0$.\footnote{An interesting behavior was also noted recently in this limit for 2D CFTs in \cite{Bachas:2021fqo}.} In this case, the Poincar\'e angle of the domain wall where it intersects the AdS boundary approaches $-\pi/2$ in the region with AdS length $\ell_0$ and $\pi/2$ in the region with AdS length $\ell_1$. In the resulting solutions, the domain wall in the $\ell_0$ region winds around the center point (Euclidean horizon) more than once (see Figure \ref{fig:patches}, right), though the solutions are still smooth, since the horizon is not included in the geometry. For these solutions, we have that $F/c_0 > 0$. If we take
\be
\label{defKappa}
\kappa = {1 \over \ell_1} - {1 \over \ell_0} + {\epsilon \over \ell_0 - \ell_1} \; ,
\ee
we find that ${F \over c_0}$ is approximately constant as a function of $z_1/z_0$, with the value
\be
{F \over c_0} \approx {1 \over \epsilon} \left(1 - {\ell_1 \over \ell_0}\right)^3 {8 {\cal I}^4\over \pi^3} \; ,
\ee
where ${\cal I}$ is the same order one constant as before. Returning to the equation (\ref{eqn}), we see that a solution requires
\be
{1 \over \epsilon} c_0 \sim c_{3D} \; .
\ee
Thus, it appears that solutions may be possible if the interface between the CFTs corresponds to an interface tension close to the minimal value\footnote{We recall that the tension parameter is related to an interface central charge (the ``boundary $F$'' for the folded theory) and that this is conjectured to decrease under RG flows, so these small values may arise naturally.}, where we take $\kappa$ as in (\ref{defKappa}) with
\be
\epsilon \sim {c_0 \over c_{3D}} \; .
\ee
It turns out that the critical value $ \kappa = {1 \over \ell_1} - {1 \over \ell_0}$ corresponds to the BPS bound for a domain wall in supergravity \cite{Cvetic:1992bf,Bachas:2020yxv}. Thus, in modelling our setup with softly broken supersymmetry, it seems natural that the interface should be modelled holographically by a domain wall with a tension close to this BPS value.

For a given choice of the tension parameter, $F$ varies very little as a function of as a function of $z_1/z_0$ (only by a fractional amount of order $\epsilon$), so it would seem that having a solution requires some fine-tuning of the tension to lie within a narrow window. It would be interesting to understand if this also occurs naturally in our supersymmetric setup. It should be noted that in the actual models, the matter in the gravitational sector is probably more accurately described by a non-conformal quantum field theory, and this may lead to additional dependence on $z_1/z_0$ that may eliminate the need for fine-tuning. Given the field theory and string theory motivations for the existence of solutions described earlier in the paper, there is reason to believe that this may be the case.\footnote{For example, in the limit of large $z_0$ for fixed $z_1$, the supersymmetry in our setup would be restored, and in this case, the vacuum energies may be expected to cancel.}

\begin{figure}
\centering
\includegraphics[width=120mm]{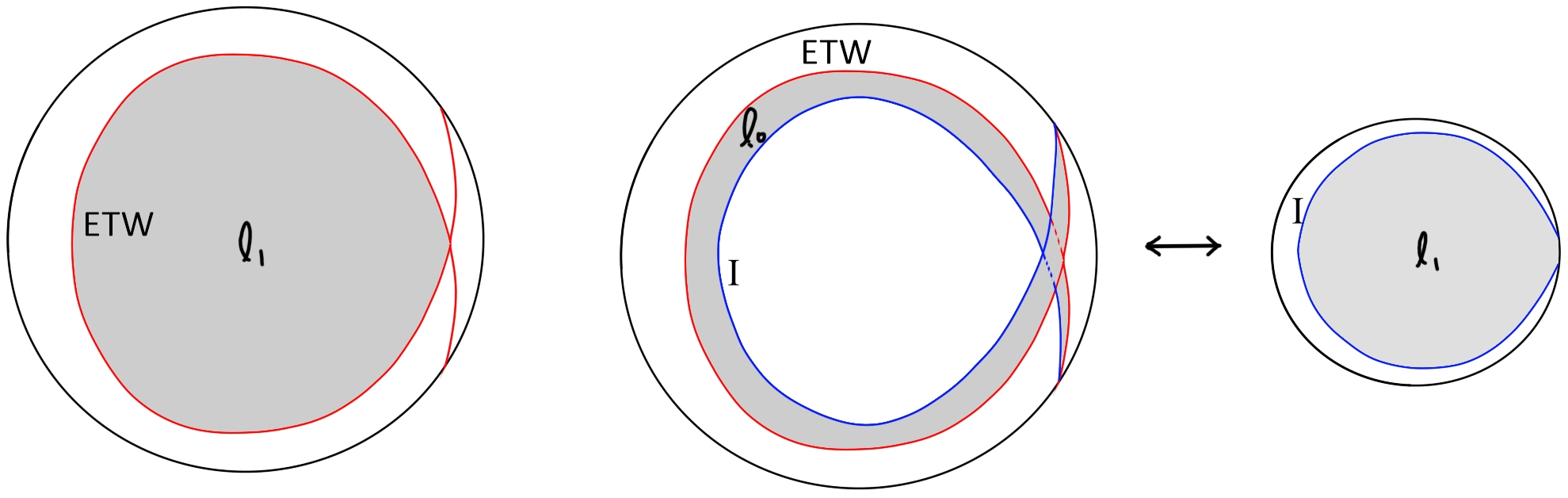}
\caption{Left: Problem with simple 5D model of gravity plus ETW brane. The ETW brane self-intersects above a tension $T_*$, below the value necessary for localization of gravity. Right: A model with an additional interface brane, which takes into account the extra matter in the gravitational theory beyond the cutoff CFT dual to the bulk. The two solutions on the right are glued along the interface brane $I$. The ETW brane and the interface are both multiply wound relative to a single copy of the Euclidean AdS/Schwarzschild geometry, so self-intersections are avoided.}
\label{fig:fiveD}
\end{figure}

Finally, we point out that the mechanism that we have found for producing large negative Casimir energies may also resolve the original puzzle in the 5D gravity description, where the desired solutions with ETW branes failed to exist because of self-intersections. In the model depicted in Figure \ref{fig:fiveD}, we have both an ETW brane and an interface brane. The extra geometrical region with AdS length $\ell_1 > \ell_0$ represents the fact that there is more matter in the effective gravity theory that in the original 4D CFT that connects the two 3D theories. Understanding in detail whether these solutions make sense may require additional input about the intersection between the interface brane and the ETW brane, but the setup appears to cure the basic pathology of a self-intersecting ETW brane.

\section{Discussion}
\label{sec:discussion}

We have presented a class of field theory constructions that may give rise to four-dimensional Euclidean wormholes, eternally traversable wormholes, and big-bang / big-crunch cosmologies in the effective description. In all cases, the gravitational theory is not purely four-dimensional, but couples to a higher-dimensional bulk which may or may not have a description in terms of classical gravity.

\subsubsection*{Comments on the effective theory}

While our immediate goal in this work is not to come up with a phenomenologically realistic model of cosmology, we mention a few interesting points related to the effective field theory description of the cosmological physics. This matter has two sectors, one coming from a cutoff version of the 4D CFT that we choose, and the other coming with the gravitational theory dual to the 3D CFTs. The gauge group in the latter sector is directly related to the global symmetry of our chosen 3D CFT. Thus, we can control the  matter that appears in the effective 4D gravity theory by choosing the 4D CFT and the global symmetry of the 3D CFTs appropriately. In the specific microscopic examples we have discussed, this global symmetry can be chosen as an arbitrary product of unitary groups.

In these and other examples where the one-boundary theory preserves supersymmetry, the matter in the sector dual to the 3D CFT should be that of some 4D gauged supergravity theory with the appropriate amount of supersymmetry (e.g. $OSp(2,2|4)$ for the microscopic theories we discussed). In the two-boundary setup relevant to cosmology, supersymmetry is broken. So we expect that some of these fields will become massive, and the effective field theory relevant to low-energy physics will not be supersymmetric. In this case, we expect that the vacuum energies of the fields would alter the cosmological constant, perhaps in a way that depends on time (in which case, the ``constant'' should be modeled using a field). In the Euclidean picture, the asymptotic value should be the negative cosmological constant associated with ETW-brane theory with unbroken supersymmetry, but this could be modified in the interior region of the ETW brane. It is interesting to ask how the construction can avoid the cosmological problem. Presumably it has to do with the very soft way that supersymmetry is broken, via incompatible boundary conditions at the boundaries in the Euclidean past and future.

\subsubsection*{Future directions}

Moving forward, it will be important to verify the existence of the proposed solutions, either from a gravity point of view (e.g. finding solutions of type IIB supergravity with the specified asymptotics, or arguing they exist), or from a field theory point of view (e.g. understanding whether the proposed microscopic theories have the suggested IR behavior and pattern of symmetry breaking).

It would be interesting to understand better the mechanism for the enhancement of negative Casimir energies presented in Section \ref{sec:casimir}. In the context of our holographic model, this can be made arbitrarily large for a CFT on a strip of fixed width by coupling the two sides via another strip CFT with smaller central charge and choosing the interface to correspond to a bulk domain wall close to a critical tension. But in microscopic examples, there is likely an upper bound on the energy, since the bulk tension corresponds to a central charge associated with the interface, and there should be some lower bound on this.

Assuming that the setup we have described is viable, it will be interesting to understand better what are the well-defined observables in the cosmological theory and how to compute these from the CFT perspective. As discussed in \cite{VanRaamsdonk:2020tlr}, these calculations may be very difficult in the Lorentzian picture where we start with a state of the auxiliary 4D CFT, since the cosmological physics happens behind a black hole horizon. But at least some observables (e.g. cosmological correlators at the time-symmetric point), seem straightforward to obtain directly from the Euclidean picture.

Finally, we discuss a connection to the physics of islands in black hole evaporation, also discussed in \cite{VanRaamsdonk:2020tlr} and in the low-dimensional models of \cite{Penington:2019kki, Dong:2020uxp, Chen:2020tes}. Following \cite{Cooper2018}, we begin by asking about the entanglement wedge for subsystems of the 4D CFT,\footnote{Here, we are talking about the picture where the Euclidean path integral is constructing some state of the Lorentzian 4D CFT on $M$.} considering a three-dimensional ball in particular in the case where the spatial geometry is $\mathbb{R}^3$. In the case where our 4D theory is holographic and we have a 5D bulk spacetime, we recall that the ETW brane lies behind a planar black hole horizon. There are two possibilities for the Ryu-Takayanagi surface \cite{Ryu:2006bv} of a ball-shaped region. We can have a surface that stays outside the black hole horizon, or we can have a surface that crosses the horizon and ends on the ETW brane. In the first case, the entropy will scale like the volume of the ball for large balls, while in the second case, the entropy will scale like the area of the ball for large balls. For large enough balls, the latter case will have lower area, and the entanglement wedge will include a portion of the ETW brane. In the case where our 4D theory is not conventionally holographic, we expect that the density matrix for a large enough ball still encodes the information about a ball-shaped region of the cosmological spacetime. In this case, this region does not have a geometrical connection to the original CFT, so it is an island in the sense of \cite{Almheiri:2019hni,Almheiri:2019yqk,Hartman:2020khs}. Thus, we expect that any cosmological spacetime with an underlying description as in this paper should have islands. In a recent paper \cite{Hartman:2020khs}, Hartman et. al. analyzed the conditions under which islands can exist in cosmological spacetimes. They found that generically, sufficiently large ball-shaped regions of radiation dominated FRW big-bang / big-crunch universes with negative cosmological constant contain ball-shaped regions satisfying the conditions to be islands. A possible explanation for the observations of \cite{Hartman:2020khs} is that the underlying microscopic description of these big-bang / big-crunch cosmologies is always similar to the one described in this paper.

\section*{Acknowledgements}

We are grateful to Juan Maldacena for questions and comments which prompted this work and guided several parts of this investigation. We also thank Nima Arkani-Hamed, Costas Bachas, Ben Freivogel, Andreas Karch, Henry Lin, Emil Martinec, Mukund Rangamani, Brian Swingle and the string theory group at UBC for useful comments and discussion. This work is supported by the Simons Foundation via the It From Qubit Collaboration and a Simons Investigator Award and by the Natural Sciences and Engineering Research Council of Canada.

\appendix

\section{Type IIB Supergravity solutions for ${\cal N}=4$ SYM theory coupled to a 3D SCFT}

\label{app:microscopic}

In this appendix, we briefly recall the solutions of \cite{DHoker:2007zhm, DHoker:2007hhe, Aharony:2011yc} corresponding to ${\cal N}=4$ SYM theory with half-supersymmetric boundary conditions. The metric is given as
\begin{equation}
    ds^{2} = f_{4}^{2} ds_{\textnormal{AdS}_{4}}^{2} + f_{1}^{2} ds_{S_{1}^{2}}^{2} + f_{2}^{2} ds_{S_{2}^{2}}^{2} +  4 \rho^{2} | dw|^{2} \: ,
\end{equation}
where $f_{1}, f_{2}, f_{4}$, and $\rho$ are real-valued functions of the complex coordinate $w = x + iy = r e^{i \theta}$, which we take to be restricted to the first quadrant $0 < \theta < \Pi/2$. We also have non-trivial dilaton, NS-NS and R-R three-form fields, and five form fields.

The explicit form of the metric and other fields may be expressed in terms of a pair $h_{1}, h_{2}$ of real harmonic functions. In terms of these, the Einstein-frame metric functions may be expressed as
\begin{equation}
    \rho^{2} = e^{- \frac{\Phi}{2}} \frac{\sqrt{ - N_{2} W}}{h_{1} h_{2}} \: , \quad f_{1}^{2} = 2 e^{\frac{\Phi}{2}} h_{1}^{2} \sqrt{ - \frac{W}{N_{1}}} \: , \quad f_{2}^{2} = 2 e^{- \frac{\Phi}{2}} h_{2}^{2} \sqrt{ - \frac{W}{N_{2}}} \: , \quad f_{4}^{2} = 2 e^{- \frac{\Phi}{2}} \sqrt{ - \frac{N_{2}}{W}} \: ,
\end{equation}
where
\begin{equation}
    e^{2 \Phi} = e^{4 \phi} = \frac{N_{2}}{N_{1}} \: ,
\end{equation}
is the dilaton field and
\begin{equation}
    \begin{split}
        W & \equiv \partial_{w} h_{1} \partial_{\bar{w}} h_{2} + \partial_{w} h_{2} \partial_{\bar{w}} h_{1} \: , \qquad X \equiv i \left( \partial_{w} h_{1} \partial_{\bar{w}} h_{2} - \partial_{w} h_{2} \partial_{\bar{w}} h_{1} \right) \: , \\
        N_{1} & \equiv 2 h_{1} h_{2} | \partial_{w} h_{1}|^{2} - h_{1}^{2} W \: , \quad \quad \: N_{2} = 2 h_{1} h_{2} | \partial_{w} h_{2}|^{2} - h_{2}^{2} W \: .
    \end{split}
\end{equation}
Explicit expressions for the other fields may be found in the references.

In general, we have a local solution for arbitrary harmonic functions $h_1,h_2$, but to obtain a global solution without singularities, we have additional constraints, e.g. that the poles must lie on the axes.

As an example, we can describe $AdS^5 \times S^5$ with the choice
\bea
\label{h12AdS}
h_1 = {L^2 \over 4}  \cos(\theta) ({r \over r_0} + {r_0 \over r}) \: , \qquad h_2 = {L^2 \over 4} \sin(\theta) ({r \over r_0} + {r_0 \over r})
\eea
Here, the codimension-one slices of the spacetime corresponding to a fixed $r$ correspond to $AdS^4 \times S^5$ slices of $AdS^5 \times S^5$ (as in Figure \ref{fig:asymptotic}); $r=r_0$ corresponds to the vertical slice. The $S^5$ arises from the angular coordinate $\theta$ and the two $S^2$s, which contract to zero on the $x$ and $y$ axes respectively.

The general solutions corresponding to ${\cal N}=4$ SYM theory on a half-space correspond to the choice\footnote{For this choice, the asymptotic value of the dilaton field has been set to zero, but we can use the symmetry $\phi \to \phi + \phi_0$, $B_2 \to e^{\phi_0} B_2$, $C_2 \to e^{- \phi_0}C_2$ to restore more general values.}
\beas
\label{gensol}
h_1 &=&{\pi  \over 2} x + {1 \over 4} \sum_A \ln \left( {(x + l_A)^2 + y^2 \over (x-l_A)^2 + y^2} \right)  \cr
h_2 &=& {\pi  \over 2}y  + {1 \over 4} \sum_A \ln \left( {x^2 + (y + k_A)^2 \over x^2 + (y-k_A)^2} \right)  \; .
\eeas
where we are choosing units with $\ell_s = 1$ \cite{Aharony:2011yc,VanRaamsdonk:2020djx}. As described in \cite{Assel:2011xz,Aharony:2011yc}, the singularities at $x = l_A,y=0$ corresponds to D5-brane throats, where the number of units $N_{D5}^A$ of D5-brane flux associated to a throat is the multiplicity of $l_A$ in the sum. Similarly, the singularities at $y = k_A, x=0$ corresponds to NS5-brane throats, where the multiplicity of the singularity $k_A$ in the sum gives the number of units $N_{NS5}^B$ of NS5-brane flux. From the five-form fluxes in the solution, \cite{Aharony:2011yc} found that the number of units of five-form flux (the flux associated with D3-branes) per fivebrane coming from the D5-branes in the $A$th stack and the NS5-branes the $B$th stack are
\bea
n_{D3}^A &=&  l_A - {2 \over \pi} \sum_B \arctan{k_B \over l_A} \cr
n_{D3}^B &=&  k_B + {2 \over \pi} \sum_A \arctan{k_B \over l_A}
\eea
Thus, microscopic solutions with properly quantized fluxes are obtained by choosing positive $k_A$ and $l_A$ (including their multiplicities) so that $n_{D3}^A$, and $n_{D3}^B$ are integers ($n_{D3}^A$ can be negative). These integer parameters are directly related to the parameters which specify the underlying field theory \cite{Assel:2011xz,Aharony:2011yc}. For example, the total amount of D3-brane flux, which corresponds to the rank $N$ of the $U(N)$ ${\cal N}=4$ SYM Theory gauge group is
\be
\label{defN}
N = \sum_A l_A + \sum_B k_B \; .
\ee
Note that the $l_A$s and $k_A$s appearing in these sums may appear with some multiplicity.

Ignoring these quantization conditions, we note that in the limit where $l_A$ and $k_A$ are taken to be small with $\sum_A l_A + \sum_B k_B$ fixed, the harmonic functions approach those corresponding to $AdS^5 \times S^5$. Thus, the full 10D supergravity solutions also approach $AdS^5 \times S^5$. In the ETW brane picture, we can say that the ETW brane angle goes to $\Pi/2$ in this limit so that we recover all of $AdS^5 \times S^5$. However, in the microscopic theory, we cannot take $l_A$ and $k_A$ arbitrarily small and still satisfy the quantization conditions.

\section{Probe D5-brane solutions}
\label{app:D5probe}

In this section, we review explicit solutions for probe D5-branes in an $AdS^5 \times S^5$ background. We use Poincar\'e coordinates for the AdS, where the metric is
\be
ds^2 = {L^2 \over z^2} (dz^2 + d \tau^2 + d \vec{y}^2)
\ee
and describe the $S^5$ by a metric
\be
ds^2 = d \psi^2 + \cos^2 \psi (d \theta^2 + \sin^2 \theta d \phi^2) + \sin^2 \psi (d \eta^2 + \sin^2 \eta d \chi^2) \; .
\ee
For a single D5-brane defect at $x = 0$ in the field theory, the D5-brane worldvolume is described by the hypersurface $\tau = \psi = 0$, filling the $z,\vec{y},\theta,$ and $\phi$ coordinates.

For parallel D5-brane defects with the same orientation, we have two probe branes at positions $x_1,x_2$. We note that the proper distance between the branes becomes small for large $z$.

For D5-branes with the opposite orientation, we also have this solution, but it is unstable. The least energy solution is a connected brane solution, studied in \cite{Antonyan:2006pg}, in which the D5-brane has some trajectory $z(\tau)$ (but still fills the $\vec{y}$ directions). On the sphere, the brane lives at $\psi = 0$ and fills the $\theta$ and $\phi$ directions.

\begin{figure}
\centering
\includegraphics[width=150mm]{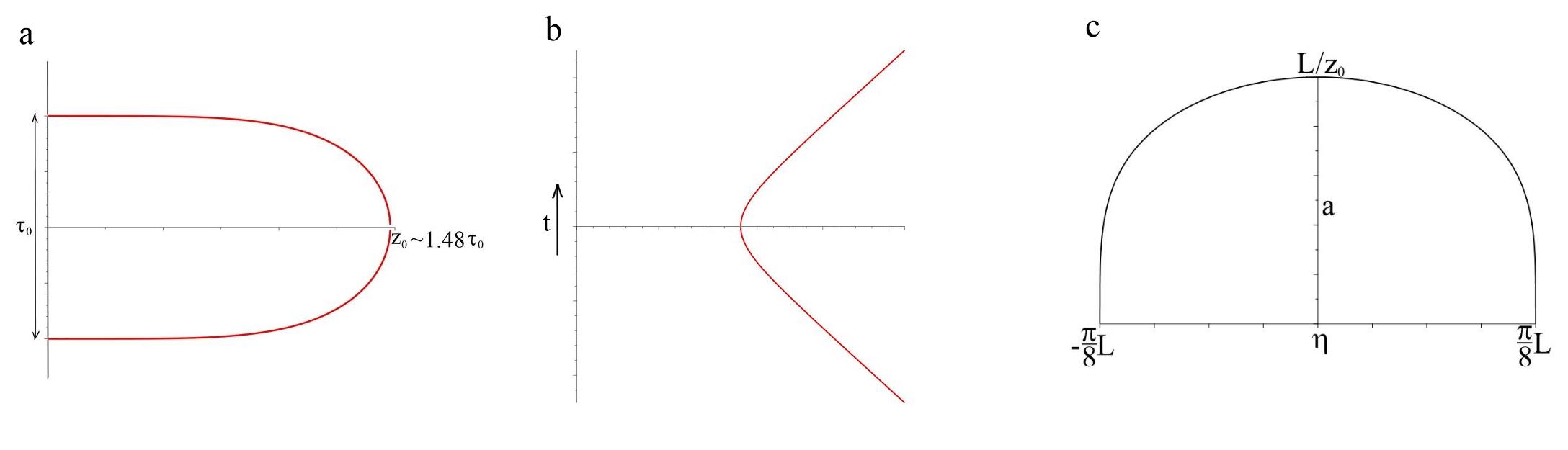}
\caption{(a) Trajectory of D5-brane probe in Poincar\'e coordinates. (b) Lorentzian trajectory of D5-brane probe in Poincar\'e coordinates. (c) Evolution of the scale factor for the worldvolume D5-brane metric in FLRW coordiantes.}
\label{fig:probecombo}
\end{figure}

The induced metric on the brane is
\be
ds^2 = {L^2 \over z^2} ((z')^2 + 1) d \tau^2 + d \vec{y}^2) + L^2 d \Omega_2^2
\ee
so the brane action gives
\be
S \sim \int d \tau \left({L \over z} \right)^4 \sqrt{1 +(z')^2} \; .
\ee
The action does not depend explicitly on $\tau$, so we find
\be
{1 \over z^4  \sqrt{1 +(z')^2}} = {1 \over z_0^4}
\ee
where $z_0$ is the maximum $z$ coordinate at the turning point on the brane. Solving, we obtain
\be
\tau(z) = z_0 \left[{\sqrt{\pi} \Gamma \left({5 \over 8} \right) \over \Gamma \left({1 \over 8} \right)} - {1 \over 5} \left({z \over z_0} \right)^5 {}_2 F_1 \left({1 \over 2},{5 \over 8} ; {13 \over 8} ; \left({z \over z_0} \right)^8 \right)\right] \; .
\ee
The parameter $z_0$ is related to the defect separation $\tau_0$ by
\be
z_0 = {\Gamma \left({1 \over 8} \right) \over 2 \sqrt{\pi} \Gamma \left({5 \over 8} \right)} \; .
\ee
This brane trajectory is plotted in Figure \ref{fig:probecombo}a.

More generally, we can consider the case with a small number $n$ of D5-branes where a small number $k$ of the D3-branes terminate on the D5s \cite{Myers:2008me, Davis:2011am,Grignani:2014vaa}. In this case, the field theory description is an interface theory where we have gauge group $U(N)$ between the defects and $U(N-k)$ outside the defects. The probe D5-branes now include $k$ units of magnetic flux on the $S^2$, and this induces $k$ units of D3-brane charge from the $\int C_4 \wedge F$ term in the D5-brane action.

Here, the relevant terms in the D5-brane action are
\be
S = -T_5 \int d^6 \sigma \sqrt{- {\rm det} (g_{ab} + 2 \pi \ell_s^2 F_{ab})} - T_5 \int 2 \pi \ell_s^2 F \wedge C_4 \; .
\ee
With our ansatz, the action governing the trajectory $z(\tau)$ becomes \cite{Myers:2008me}
\be
S \sim \int d \tau z^{-4} \left\{ \sqrt{(1 +(z')^2)(1 + f^2)} \pm f \right\}\; ,
\ee
where $f = \pi q /(\sqrt{\lambda} n)$. In this case, the probe branes tilt outward as they enter the bulk, at an angle $\tan \theta = f$ from the radial direction in Poincar\'e coordinates, but we have a connected solution provided that $f < C$ where $C \approx 0.357$, i.e. provided the number of D3-branes per fivebrane is not too large. In our full construction, we want a relatively small number of D3-branes and a large number of fivebranes, so this condition should be satisfied. However, we should not conclude too much from the probe analysis.

\subsubsection*{Lorentzian probe brane solutions}

As an example of the connection to cosmology, we can determine the FRW spacetime associated to the probe brane trajectory (without D3 charge), though here, gravity does not localize to the brane. In the analytically continued case, the brane exists the Poincar\'e horizon, reaches some minimum $z$ value $z_0$ and then falls back into the horizon.

The trajectory satisfies
\be
{1 \over z^4  \sqrt{1 -(\dot{z})^2}} = {1 \over z_0^4} \; .
\ee
We can write the solution as
\be
t(z) =  z \; {}_2 F_1 \left(-{1 \over 2},{1 \over 2} ; {7 \over 8} ; \left({z_0 \over z} \right)^8 \right) -{z_0 \over \sqrt{\pi}} \Gamma \left({5 \over 8} \right)\Gamma \left({7 \over 8} \right) \sin \left({3 \pi \over 8} \right) \; .
\ee
This is displayed in Figure \ref{fig:probecombo}b.

The worldvolume metric can be written most simply using $z$ to parameterize the time direction. This gives (for the metric in the $t>0$ region
\be
ds^2 = {L^2 \over z^2} \left[-{z_0^8 dz^2 \over z^8 - z_0^8} + d \vec{y}^2\right]
\ee
In order to see the evolution of the scale factor, we can convert to standard flat FLRW coordinates
\be
-d \eta ^2 + a^2(\eta) d \vec{y}^2 \; .
\ee
We have
\be
a(\eta) = {L \over z(\eta)} \qquad \qquad ds = {L \over z} {z_0^4 \over \sqrt{z^8 - z_0^8}} dz
\ee
which gives
\be
a(\eta) = {L \over z_0} \cos^{1 \over 4} \left({4 \eta \over L}\right)
\ee
This is plotted in Figure \ref{fig:probecombo}c.

\subsection{Probe $\bar{D5}$-brane in the background of a single-boundary solution.}
\label{app:boundprobe}

\begin{figure}
\centering
\includegraphics[width=80mm]{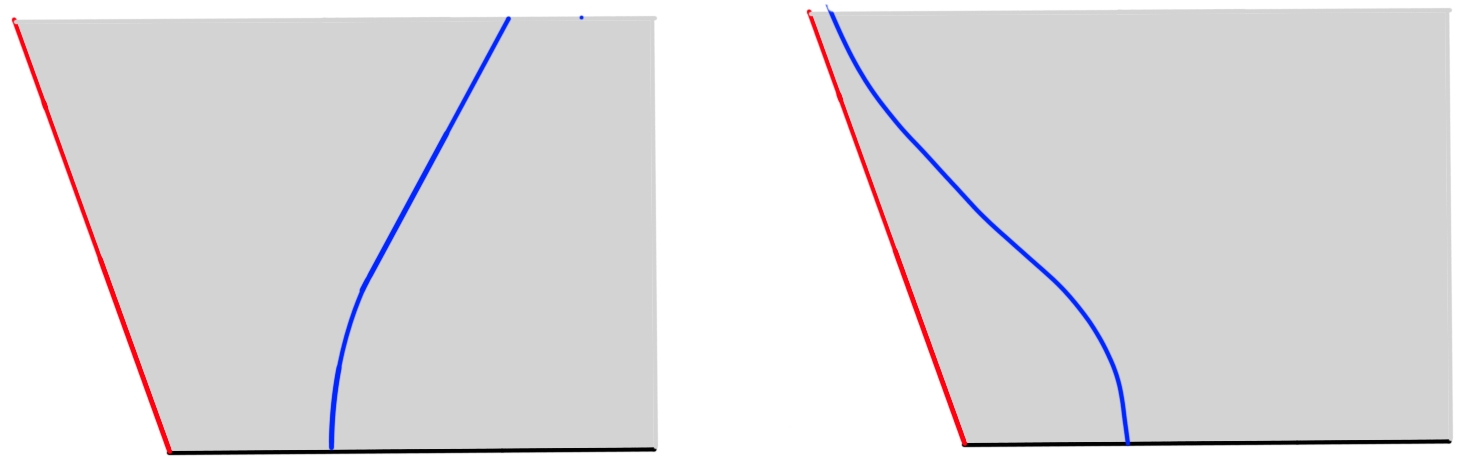}
\caption{Probe brane in the background dual to ${\cal N}=4$ SYM with one SUSY boundary. Left: a SUSY-preserving D5-brane probe remains separated from the ETW brane (left). Right: a SUSY-breaking $\bar{D5}$ is drawn towards the ETW brane, though we can also have solutions of the first type in cases where the full solution is very close to $AdS^5 \times S^5$. }
\label{fig:boundprobe}
\end{figure}

In this section, we consider a probe D5-brane in the general half-supersymmetric supergravity backgrounds reviewed in the previous section. The brane action is the sum of Born-Infeld and Wess-Zumino terms, given in Einstein frame by
\be
S_{BI} = -T_5 \int d^6 \sigma e^{\phi \over 2} \sqrt{- {\rm det} (g_{ab} + e^{-{\phi \over 2}} (B_{ab} + 2 \pi \alpha' F_{ab}))}
\ee
\be
S_{WZ} = -T_5 \int e^{2 \pi \alpha' F + B} \wedge \sum C \; .
\ee
As for the $AdS^5 \times S^5$ solutions, we have that the D5-branes live at $\theta = 0$, wrapping the first $S^2$. They are described by some trajectory $r(u)$, where $u$ is the Poincar\'e radial direction in $AdS^4$ and $r$ is the radial coordinate on the quadrant. The brane is stretched in the other three directions of $AdS^4$.

The worldvolume gauge field can be consistently set to zero.\footnote{It's possible to consider solutions with world-volume flux, but these correspond to fivebranes carrying additional D3-brane charge}. Using the form of the supergravity solution, we then find a Lagrangian density
\be
{\cal L} =  {A(r) \over u^4} \sqrt{1 + B(r) u^2 \left({du \over dr}\right)^2} + {K(r) \over u^4}
\ee
where\footnote{We find that $C^{(6)}$ vanishes for $\theta=0$, so the Wess-Zumino term only receives a contribution from $B \wedge C^{(4)}$.}
\beas
A(r) = e^{\phi \over 2} f_4^3 \sqrt{f_1^4 + e^{-\phi} B_{45}^2} \cr
B(r) = 4 { \rho^2 \over f_4^2} \cr
K(r) = \pm B_{45} C_{0123}
\eeas
where the two possible signs in $K$ correspond to a D5-brane probe or and anti D5-brane probe. Redefining $u = exp(x)$, we obtain equations of motion
\be
\label{ProbeEOM}
ABr'' - {d K \over dr}(1 + B (r')^2)^{3 \over 2} - {dA \over dr} - 4 A B^2 (r')^3 + {1 \over 2} (r')^2 (A {d B \over d r} - 2 {d A \over dr} B) - 4 A B r' = 0
\ee
We want to consider a probe brane starting at $r=\infty$ at some fixed $x$. Solutions that return to $r=\infty$ at some fixed $x$ can be ignored since these correspond to having additional defects. We cannot have a solution that goes to $r = \infty$ for $x \to \infty$ since the solution approaches $AdS^5 \times S^5$ for large $r$ and we do not have such solutions in this case. Thus, we have two possibilities: the solution approaches some finite $r$ for $x \to \infty$ in the region away from the ETW brane, or the brane is drawn toward the ETW brane (specifically to one of the D5-brane singularities).

For the first type of solution, we have $r' \to 0$ for large $x$, so (\ref{ProbeEOM}) gives
\be
r'' = {1 \over AB}\left({dA \over dr} + {dK \over dr} \right) \; .
\ee
Thus, we can have solutions where the probe brane is not drawn into the ETW brane if and only if $A+K$ has a extremum. As an example, for $AdS^5 \times S^5$, we have
\be
A + K  \propto \left({r \over r_0} + {r_0 \over r} \right)^4
\ee
so we have a minimum at $r=r_0$ which corresponds to the vertical slice in Poincar\'e coordinates.

We have investigated the probe brane configurations for various parameter values. With a SUSY-preserving D5 orientation, it appears that solutions of the first type always exist. On the other hand, replacing the D5 with an anti-D5, we find that such solutions do not exist in many cases so the anti-D5 brane is necessarily drawn in to the ETW brane. This is a probe version of the situation where two ETW branes from two boundaries reconnect in the bulk. On the other hand, we have solutions that are arbitrarily close to $AdS^5 \times S^5$. Since $A + K$ has a minimum for $AdS^5 \times S^5$ it will continue to have a minimum in solutions that are very close to $AdS^5 \times S^5$, for either sign of $K$, but these solutions correspond to boundary conditions involving a very large number of D5-branes. So in this case, considering a single D5-brane probe may not give us useful insight about the physics of adding a second boundary that would involve a very large number of anti-branes.

\section{Holographic model for conformal interfaces}
\label{app:interface}

The details of our holographic model for conformal interface theories can be found in \cite{Simidzija:2020ukv}. We include here the formulae for $z_0/\beta_0$ and $z_1/\beta_1$, the fraction of the boundary of Euclidean global AdS spacetime covered by the spacetime regions associated with CFT${}_0$ and CFT${}_1$ respectively.  We have
\beas
{z_0 \over \beta_0} &=& -{2 \mu_0^{3 \over 4} \over \pi \ell_0^{3 \over 2}} \int_{r_0}^\infty {dr (f_0 - f_1 + \kappa^2 r^2) \over 2 \kappa r f_0 \sqrt{V}} \cr
{z_1 \over \beta_1} &=& 1 - {2 \mu_1^{3 \over 4} \over \pi \ell_1^{3 \over 2}} \int_{r_0}^\infty {dr (f_1 - f_0 + \kappa^2 r^2) \over 2 f_1 \kappa r \sqrt{V}}
\eeas
where
\be
f_i = {r^2 \over \ell_i^2} - {\mu_i \over r^2}  \qquad V = f_1 - \left({f_2 - f_1 - \kappa^2 r^2 \over 2 \kappa r} \right)^2
\ee
and $r_0$ is the largest value of $r$ for which $V(r_0)=0$. These formulae are valid in cases such as Figure \ref{fig:patches}b where the CFT${}_0$ side does not include the Euclidean horizon but the CFT${}_1$ side does. We have the restriction that $z_1/\beta_1 < 1$ to have a sensible solution, but $z_0/\beta_0$ can be greater than 1 in the case where we have multiple wound solutions as in Figure \ref{fig:patches}b.

\section{Effective field theory description with conformal anomaly}
\label{app:anomaly}

In the analysis of section, we have ignored the conformal anomaly. However, if we assume that the CFT${}_0$ matter theory is holographic with an Einstein gravity dual, it is possible to explicitly take the conformal anomaly into account.

The contribution to the stress tensor from the conformal anomaly can be determined via a gravity calculation to be
\be
T_{zz} = - {3 \hat{c}_0 \over 16 a^2}\left({a' \over a} \right)^4
\ee
where up to a numerical factor, $c_0$ is the $a$ or $c$ type central charge of CFT${}_0$ (which are equal for a holographic CFT with Einstein gravity dual).

With this contribution, the $zz$ component of Einstein's equation gives
\be
 \left({a' \over a} \right)^2 - {a^2 \over L_{AdS}^2} = - {8 \pi G \over z_0^4 } F {1 \over a^2} - { \pi G c_0 \over 2 a^2}\left({a' \over a} \right)^4 \; ,
\ee
or
\be
{da \over  dz} = {a^2 \over \sqrt{\pi G c_0}} \sqrt{\sqrt{1 + {2 \pi G c_0 \over L^2} - {16 \pi^2 G^2 F c_0 \over z_0^4 a^4}}-1}
\ee
The minimum value of $a$ is still
\be
a_{min} = {1 \over z_0} (8 \pi G F L^2)^{1 \over 4} \; .
\ee
Integrating from this minimum radius (which occurs at $z=0$) to the asymptotically AdS boundary at $z=z_0/2$ and rearranging things, we get
\be
\label{eqn2}
F \left( {z_1 \over z_0} \right) = c_{3D} Q\left({c_0 \over c_{3D}}\right)
\ee
where we have taken $c_{3D} = L^2/(2 \pi G)$ and
\be
Q(\epsilon) = \left[ \int_0^1 {\sqrt{\epsilon}dy \over \sqrt{\sqrt{1 + \epsilon(1 - y^4)}-1}}\right]^4
\ee
We can check that in the limit $c_0/c_{3D} \to 0$,$Q(c_0/c_{3D})$ approaches a constant to give the same equation as before. For small $c_0/c_{3D}$ (as we expect), we get a slightly different number on the right hand side of (\ref{eqn2}).

\subsection{Lorentzian solutions}

Including the effects of the conformal anomaly does have some interesting consequences for the Lorentzian solutions, namely the existence of a minimum scale factor and/or maximum initial energy density.

If we work in standard FRW coordinates,
\be
\label{FRW}
-dt^2 + A^2(t) d^3 x
\ee
the Einstein equation leads to the Friedmann equation
\be
 \left({\dot{A} \over A} \right)^2 + {1 \over L_{AdS}^2} = {\pi G c_0 \over 2} \left[\left({\dot{A} \over A} \right)^4 +  {16 F \over z_0^4 c_0 }  {1 \over A^4} \right] \; .
\ee
Defining
\be
\hat{A} = z_0 \left({c_0 \over 16 F}\right)^{1 \over 4} {1 \over \sqrt{\pi G c_0}} A \qquad \hat{t} =  {t \over \sqrt{\pi G c_0}} \qquad \hat{L} =  {L \over \sqrt{\pi G c_0}} \sim {c_{3D} \over c_0} \; ,
\ee
the equation simplifies to
\be
\label{Feq}
\left({d \hat{A} \over d \hat{t}}\right)^2 = \hat{A}^2 \pm \sqrt{\left({2 \over \hat{L}^2} + 1\right) \hat{A}^4 - 1}
\ee
The two signs here correspond to two branches of solutions, one giving expanding and contracting solutions ($-$ sign) and the other giving rise to inflating spacetimes.\footnote{This mechanism for inflation is closely related to Starobinsky's original model for inflation \cite{Starobinsky:1980te}.} We are interested in the former.

For either branch of solutions, we have a minimum possible value of $\hat{A}$, where the expression inside the square root vanishes,
\be
\hat{A}_{min} = \left({2 \over \hat{L}^2} + 1\right)^{-{1 \over 4}} \; .
\ee
Presumably, this is where the semiclassical approximation is breaking down. The energy density at this point is
\be
T_{00} \sim {1 \over c_0 G^2}
\ee
dominated by the conformal anomaly contribution.

The explicit solutions for $a(t)$ can be obtained by integrating (\ref{Feq}); the scale factor expands from its minimum value to a maximum
\be
\hat{A}_{max} = \left({\hat{L}^2 \over 2} \right)^{1 \over 4}
\ee
before contracting. We have
\be
{\hat{A}_{max} \over \hat{A}_{min}} = \left({\hat{L}^2 \over 2} + 1\right)^{1 \over 4} \sim \left(c_{3D} \over c_0 \right)^{1 \over 4}
\ee
so we have a large amount of expansion when $c_{3D} \gg c_0$, i.e. the number of local degrees of freedom in the original holographic 3D CFT is much larger than the number of matter degrees of freedom in the dual theory.

\bibliographystyle{jhep}
\bibliography{references}

\end{document}